\documentclass[11pt,a4paper]{article}
\pdfoutput=1
\usepackage{jcappub}

\title{Exact spherically-symmetric inhomogeneous model with \boldmath $n$ perfect fluids}

\author[a,b,c]{Valerio Marra}
\author[b,c]{and Mikko P\"a\"akk\"onen}

\affiliation[a]{Institut f\"ur Theoretische Physik, Universit\"at Heidelberg, Philosophenweg 16,\\
69120 Heidelberg, Germany}

\affiliation[b]{Department of Physics, PL 35 (YFL), 40014 University of Jyv\"askyl\"a, Finland}

\affiliation[c]{Helsinki Institute of Physics, PL 64, 00014 University of Helsinki, Finland}

\emailAdd{valerio.marra@me.com}
\emailAdd{mikko.u.paakkonen@jyu.fi}

\abstract{
We present the exact equations governing the dynamics of a spherically-symmetric inhomogeneous model with $n$ decoupled and non-comoving perfect fluids.
Thanks to the use of physically meaningful quantities we write the set of $3+2n$ equations in a concise and transparent way.
The $n$~perfect fluids can have general equations of state, thus making the model extremely flexible to study a large variety of cosmological and astrophysical problems.
As applications we consider a model sourced by two non-comoving dust components and a cosmological constant, and a model featuring dust and a dark energy component with negligible speed of sound.
}


\arxivnumber{1105.6099}

\begin{document}
\maketitle

%
%

\section{Introduction} \label{intro}

Spherically-symmetric models are of interest in the study of the nonlinear inhomogeneities of the universe for two reasons.
First, spherical symmetry is often a reasonable working approximation for the inhomogeneities in the real universe.
Second, spherical symmetry allows to exactly solve the Einstein's equations, a highly nontrivial task.
Recent research has focused mainly on the Lema\^{i}tre model \cite{Lemaitre:1933gd} (see \cite{Bolejko:2005tk, Lasky:2006mg, Alfedeel:2009ef, Lasky:2010vn} for recent contributions) which describes the dynamics of a spherically-symmetric perfect fluid, and great attention has received its pressureless limit, usually named the Lema\^{i}tre-Tolman-Bondi (LTB) model~\cite{Lemaitre:1933gd,Tolman:1934za,Bondi:1947av}, which can also include a nonvanishing cosmological constant.
LTB metrics have been used to describe the large-scale inhomogeneities of the late universe as, for example, in Swiss-cheese~\cite{Biswas:2007gi, Marra:2007pm, Marra:2007gc, Brouzakis:2007zi, Marra:2008sy, Vanderveld:2008vi, Valkenburg:2009iw, Szybka:2010ky}, void~\cite{Celerier:1999hp,Tomita:1999qn, Moffat:1994qy, Alnes:2005rw, Chung:2006xh, Enqvist:2006cg, Tanimoto:2007dq, Alexander:2007xx, GarciaBellido:2008nz, Yoo:2008su, Zibin:2008vk, Kainulainen:2009sx, February:2009pv, Sollerman:2009yu, Kolb:2009hn, Dunsby:2010ts, Yoo:2010qy, Biswas:2010xm, Clarkson:2010ej, Moss:2010jx,Marra:2010pg} and inhomogeneous~\cite{Mustapha:1998jb,Iguchi:2001sq,Celerier:2009sv} models for apparent acceleration (see \cite{Marra:2011ct, Bolejko:2011jc} for recent reviews).
Besides being useful in understanding the role of voids in the process of structure formation~\cite{Sheth:2003py, Bolejko:2004vb}, the LTB model is also suitable to study the dynamics of spherical collapse in an expanding universe, black-hole formation included~\cite{Joshi:1993zg, Krasinski:2003cx, Firouzjaee:2008gs, Mimoso:2009wj, Valkenburg:2011tm}.

In the present paper we extend the Lema\^{i}tre model to the case of $n$ decoupled and non-comoving perfect fluids with general equations of state.
Our main result is the set of $3+2n$ exact equations governing the dynamics of the model, which we write in terms of physically meaningful quantities and express in a concise and transparent way.
Many are the possible applications, both at early and late times.
At late time, one can study the evolution of overdensities and underdensities in a universe where dark energy is not the cosmological constant and, in particular, can be inhomogeneous (see, for example, \cite{Bertacca:2010ct,Li:2011sd} and references therein).
This possibility may also induce an inhomogeneous variation of fundamental constants, as for instance the fine structure constant if a coupling between the dark energy and the electromagnetic field is allowed (see, for example, \cite{Bekenstein:1982eu, Olive:2001vz, Marra:2005yt, Copeland:2006wr, Webb:2010hc}).
Moreover, dark matter and baryons can be described as two separate fluids, the latter possibly featuring pressure.
As stressed in~\cite{Lasky:2010vn}, the introduction of non-dust equations of state can have a non-negligible effect on the cosmological models and this should be taken into consideration while interpreting cosmological datasets.
This is, however, a non-trivial task due to the increased number of free parameters, and such difficulties increase when one considers $n$ fluid components with the consequence that more astronomical and astrophysical input is needed to constrain the available parameter space.
Finally, at early times the contribution of radiation can be included, which may be relevant both for the understanding of the evolution of the standard post-inflation inhomogeneities and for the correct modeling of the very large underdensities typical of void models~\cite{Clarkson:2010ej}.

This paper is organized as follows.
In Section \ref{model} we will go through all the details of our formalism, and in Sections \ref{colla} and \ref{cde} we will present the numerical results for the case of two non-comoving dust components in a flat $\Lambda$CDM universe and for spherical collapse in the presence of dark energy with negligible speed of sound, respectively.
Then in Section \ref{litera} we will compare our findings to previous work dealing with exact solutions.
Finally we will give our conclusions in Section \ref{conclusions}.
In Appendix \ref{exten} we discuss the expansion tensor and in Appendix \ref{pLTB} the case of the Lema\^{i}tre metric, possibly in a non-comoving frame.
The definitions of the functions used to model the numerical example of Section \ref{colla} are given in Appendix \ref{wdn}.

\section{The model} \label{model}

\subsection{Metric and Einstein tensor}

A spherically-symmetric metric can be written as
\begin{equation} \label{metric}
   \textrm{d}s^2=-e^{2\lambda}\textrm{d}t^2+\frac{Y'^2}{1+2E}\textrm{d}r^2+Y^2\textrm{d}\Omega^{2} \,,
\end{equation}
where the lapse function $e^\lambda$, the scale function $Y$ and the curvature function $E$ depend on the coordinate time $t$ and coordinate radius $r$, $\textrm{d}\Omega^{2}= \textrm{d} \theta^{2} + \sin^{2}\theta \, \textrm{d} \phi^{2}$ and we have set $c=1$.
A prime denotes partial derivation with respect to the coordinate radius $r$, whereas a dot denotes partial derivation with respect to the coordinate time $t$.
Comma and semicolon signs will {\em not} denote derivatives.
We will use a reference frame comoving with the arbitrary four-velocity field $u_{\rm rf}^{\alpha}$, which has components $u_{\rm rf}^{\alpha}=(e^{-\lambda},\;0,\;0,\;0)$.
Moreover,  we assume that $Y'\neq 0$ and $E>-1/2$ in order to always have a regular $g_{rr}>0$.
We will discuss more general evolutions and shell crossing in forthcoming work.

The nontrivial components of the Einstein tensor $G_{\alpha\beta}\equiv R_{\alpha\beta}-\frac{1}{2}R \, g_{\alpha\beta}$ for the metric~(\ref{metric}) are then:
\begin{eqnarray*}
   G_{tt}           & = &  e^{2\lambda} \left[-2 {(E Y)' \over Y^2Y'}  + H_A \Big( H_A +  2H_R    \Big)    \right] , \qquad  \quad G_{tr} = 2 e^{ \lambda} \left(  \, \lambda'   H_A   -   E_{d1}  \frac{Y'}{Y}   \right) , \\
   G_{rr}           & = &  -\frac{Y'^2}{1+2E}   \left(   H_A^{2} +  2 A_{A}  -2  {1+2E \over Y'Y} \lambda' -{2E \over Y^{2}}    \right) \,, \\
   G_{\theta\theta} & = & G_{\phi \phi}/ \sin^{2}\theta = - Y^{2}   \left(  A_{R} + A_{A} + H_{R} H_{A} -{E' \over Y'Y} - {1+2E \over Y'^{2}} \lambda' {\mathcal F}  \right)   ,  
\end{eqnarray*}
where we have defined radial and angular expansion rates:
\begin{eqnarray}
H_R&=& {\sqrt{1+2E} \over Y'} \, {d {Y' \over \sqrt{1+2E}} \over d\tau_{\rm rf}}= {e^{-  \lambda}\dot{Y}' \over Y'} - E_{d1}   \label{hubbleR} \,, \\
H_A&=&  {1 \over Y} \, {d  Y \over d\tau_{\rm rf}} = {e^{-  \lambda} \dot{Y} \over Y} \label{hubbleA} \,,
\end{eqnarray}
and also radial and angular acceleration rates:
\begin{eqnarray}
A_R&=& {\sqrt{1+2E} \over Y'}  {d^{2} {Y' \over \sqrt{1+2E}} \over d\tau_{\rm rf}^{2}}={e^{-2 \lambda} \ddot{Y}' \over Y'} -2 E_{d1} H_{R} + E_{d1}^{2} - E_{d2} - e^{-  \lambda} \dot{\lambda}   H_R \label{accR} , \\
A_A&=&  {1 \over Y}  {d^{2}  Y \over d\tau_{\rm rf}^{2}} = {e^{-2 \lambda} \ddot{Y} \over Y} -e^{-  \lambda} \dot{\lambda}   H_A   \label{accA} \,,
\end{eqnarray}
respectively, and also the following auxiliary quantities:
\begin{equation*}
E_{d1}= {e^{-  \lambda} \dot{E} \over 1+2E} \,, \qquad
E_{d2}={e^{- 2 \lambda} \ddot{E} \over 1+2E} \,, \qquad
{\mathcal F}= \frac{Y'}{Y} - \frac{Y''}{Y'}  + {E' \over1+2E}   +\lambda'  + { \lambda'' \over \lambda' } \,.
\end{equation*}
Note that for $\lambda=0$ and $\dot E=0$ the usual LTB expressions are recovered.

In the previous equations $d/d\tau_{\rm rf}=u_{\rm rf}^{\alpha} \, \partial_{\alpha}= e^{-\lambda} \partial/\partial t$ is the derivative with respect to the proper time of the comoving observer.
As we show in Appendix \ref{exten}, the components of the acceleration of $u_{\rm rf}^{\alpha}$ are $a_{r,\, \rm rf}=\lambda'$ and $a_{t,\, \rm rf}=0$.
We see therefore that, for a geodesic reference-frame velocity field, the lapse function $\lambda$ depends only upon time and can be rescaled such that $\lambda=0$ and $d/d\tau_{\rm rf} =\partial/\partial t$.
From Eqs.~(\ref{hubbleR}-\ref{hubbleA}) it follows then that in a non-geodesic reference frame the comoving observers measure with different proper times.
Later we will identify the reference-frame velocity $u_{\rm rf}^{\alpha}$ with the velocity field of one of the fluid components, and it will turn out that the acceleration $a_{r,\, \rm rf}=\lambda'$ is sourced by pressure gradients which push the observers out of the freely-falling geodesics.

\subsection{Einstein's equations and energy-momentum tensor}

From the Einstein's equations $G_{\alpha\beta}= \kappa \, T_{\alpha\beta}$ we can obtain four independent dynamical equations.
We choose to form two of them with the following combinations:
\begin{eqnarray}
     {Y^2Y'G_{tt}-Y^2\dot{Y}G_{tr}  \over e^{2\lambda}}   &\equiv& \big(e^{-2\lambda}Y\dot{Y}^2 -2 E Y \big)' = \kappa  {Y^2 Y' T_{tt}- Y^2 \dot{Y} T_{tr}  \over e^{2\lambda}} \,,  \label{ee1}  \\
     {Y^2\dot{Y}G_{rr}-Y^2Y'G_{tr} \over -Y'^2/ (1+2E) }   &\equiv&   \big(e^{-2\lambda}Y\dot{Y}^2 -2 E Y \big)\dot{}  = \kappa    {Y^2\dot{Y}T_{rr}-Y^2Y'T_{tr} \over - Y'^2/ (1+2E) }  \,,   \label{ee2}
\end{eqnarray}
where $\kappa= 8 \pi G$ and $T^{\alpha\beta}$ is the energy-momentum tensor for an ideal fluid source comprised of $n$ decoupled components:
\begin{equation}
   T^{\alpha\beta}=\sum_{i=1}^{n} T^{\alpha\beta}_{i} \,,
\end{equation}
where the energy-momentum tensor of the $i$:th perfect fluid component is:
\begin{equation}
T^{\alpha\beta}_{i}  =  \rho_i \, u_i^{\alpha}u_i^{\beta} +  p_i \, h^{\alpha\beta}_{i}  \,,
\end{equation}
where $u^{\alpha}_{i}$, $\rho_{i}$ and $p_{i}$ are the four-velocity field, energy density and pressure, respectively, and $h^{\alpha\beta}_{i}=g^{\alpha\beta} + u_i^{\alpha}u_i^{\beta} $ is the projection tensor on the hypersurface orthogonal to $u_{i}^{\alpha}$. The isotropic pressure is related to the energy density by $p=w \, \rho$, where the equation of state parameter $w=w(t,r)$ is assumed to be a general function of $t$ and $r$ (see, for example, \cite{Grande:2011hm} for the case of a static fluid with anisotropic pressure).
In the chosen coordinate system it is
\begin{equation}
u_i^{\alpha}=\gamma_i \, (e^{-\lambda},\;v_{i,\,c},\;0,\;0) \,,
\end{equation}
where $v_{i,\,c}$ is the {\em coordinate} comoving peculiar velocity of the $i$:th component relative to the reference frame. The {\em proper} peculiar velocities and the gamma factors are instead given by
\begin{equation}
v_{i,\,p}^{2}= \frac{Y'^2}{1+2E} \, v_{i,\,c}^2
\qquad \textrm{and} \qquad
\gamma_i^{2}= {1 \over  1-v_{i,\,p}^{2}}   \,,
\end{equation}
respectively.
The covariant components of the total energy-momentum tensor are then:
\begin{eqnarray}
   T_{tt}           & = & e^{2\lambda}  \sum_{i=1}^{n} \rho_i  \left[ (1+w_i)\gamma_i^2-w_i \right]  \,,  \label{BTtt}  \\
   T_{tr}           & = & - \frac{Y'^2}{1+2E}e^{\lambda} \sum_{i=1}^{n} \rho_i (1+w_i) v_{i,\,c} \gamma_i^2    \,,  \label{BTtr} \\
   T_{rr}           & = & \frac{Y'^2}{1+2E}  \sum_{i=1}^{n} \rho_i \left[ (1+w_i)v_{i,\,p}^2\gamma_i^2 + w_i \right]  \,,  \label{BTrr} \\
   T_{\theta\theta} & = & T_{\phi \phi}/ \sin^{2}\theta   =  Y^2   \sum_{i=1}^{n} p_i  \,. \label{BTthth} 
\end{eqnarray}

As the third dynamical equation we will consider the combination coming from $-G^{t}_{t}+G^{r}_{r}+G^{\theta}_{\theta}+ G^{\phi}_{\phi}$, which gives the generalization of the acceleration equation:
\begin{equation} \label{acceq}
A \equiv A_{R} + 2 A_{A} - {\sqrt{1+2E}\over Y'} \, a_{\rm rf} \left (  \frac{Y'}{Y} +{\mathcal F} \right )= - {\kappa \over 2} \sum_{i=1}^{n} \rho_i (1 + 3 w_i ) \,,
\end{equation}
where we used the fact that the acceleration scalar of the reference-frame velocity is $a_{\rm rf}  = {\sqrt{1+2E} \over Y'} \lambda'$ (see Appendix \ref{exten}).
The term proportional to $a_{\rm rf}$ gives a ``spurious'' contribution to the acceleration and vanishes if we use a geodesic reference frame, in which the total acceleration is $A_R+ 2 A_A$, similarly to LTB models.

Finally, the last independent equation will be simply the $G_{tr}=\kappa \, T_{tr}$ component, which reads:
\begin{equation} \label{kdot1}
{ e^{-  \lambda} \dot{E}  \over 1+2E}  =  {\kappa \over 2} {Y \over \sqrt{1+2E}} \sum_{i=1}^{n} \rho_i (1+w_i) v_{i,\,p} \gamma_i^2 
+  {e^{-  \lambda} \dot Y \over \sqrt{1+2E}} \, a_{\rm rf}  \,.
\end{equation}
Eq.~(\ref{kdot1}) shows that the evolution of the curvature is due to two distinct causes.
The first is the effect of having an inhomogeneous multicomponent fluid and goes to zero in the FLRW limit where the peculiar velocities vanish or if only one fluid is present  and its reference frame is adopted.\footnote{Note, however, that the cosmological constant never sources $\dot E$.}
It is sourced by the energy flux in the radial direction: we remind indeed that the curvature function $E$ may be interpreted as the total energy of a given shell (see Eq.~(\ref{energy})).
The second contribution to $\dot E$ is due to the fact that, generally, we are using a non-geodesic reference frame ($a_{\rm rf} \neq 0$) in which the total energy of a shell $r$ is not conserved.

\subsection{Conservation equations} \label{consese}

As we are considering an ideal fluid source comprised of $n$ decoupled components, each $i$:th component will be conserved individually, i.e., $\nabla_{\alpha} T^{\alpha\beta}_{i}=0$, where $\nabla_{\alpha}$ denotes the covariant derivative.
For each $i$:th component we obtain the two following equations\footnote{The other two equations relative to the angular components identically state the spherical symmetry of the energy-momentum tensor.} where we omit the index $i$ and multiply for clarity's sake by $e^{\lambda}$ and $v_c \frac{Y'^2}{1+2E}$, respectively:
\begin{eqnarray}
e^{\lambda}  \nabla_{\alpha} T^{\alpha t} & = &  (\rho+p) \gamma      \left(  \Theta +   v_{p}  a   \right)  + v_c \gamma^{2}  (\rho+p)'   +     e^{-\lambda}  \dot{p}  v^{2}_{p} \gamma^{2}  +   e^{- \lambda}\dot{\rho}  \gamma^{2}    = 0     ,  \label{cx1} \\
\frac{v_c  Y'^2}{1+2E} \nabla_{\alpha}  T^{\alpha r} & = &        (\rho+p) \gamma      \left ( v^{2}_{p} \Theta +   v_{p}  a   \right)   +   v^{2}_{p} \gamma^{2} e^{-\lambda}  (\rho+p) \dot{}  +   p'   v_{c} \gamma^2 + v_{c}   \rho'   \gamma^{2} v_p^2  = 0  , \phantom{abc}  \label{cx2} 
\end{eqnarray}
where the expansion scalar $\Theta$ and the acceleration scalar $a$ have been calculated in Appendix \ref{exten}.
By taking the combinations $e^{\lambda} \, \nabla_{\alpha} T^{\alpha t}  - v_c  \frac{Y'^2}{1+2E}  \nabla_{\alpha} T^{\alpha r}$ and $e^{\lambda} \, \nabla_{\alpha}  T^{\alpha t} - {1 \over v_{c}}\nabla_{\alpha} T^{\alpha r}$ it is possible to obtain the relativistic energy-conservation and Euler equations, respectively, for non-comoving fluids:
\begin{eqnarray}
{d \rho \over d \tau}& = & -\Theta \, (\rho+p) \,,  \label{cons1} \\
{d p \over d \sigma} &=& -  a \,  (\rho+p)  \,, \label{cons2} 
\end{eqnarray}
where, analogously to the total derivative with respect to the proper time $\tau$, we defined the convective derivative with respect to $\sigma$ along the four-acceleration $a^{\mu}$:
\begin{eqnarray}
{d \over d\tau} &=& u^{\mu} \partial_{\mu} = \gamma e^{-\lambda} {\partial \over \partial t} + \gamma v_{c} {\partial \over \partial r} 
\,, \label{dtau} \\
{d \over d\sigma} & \equiv &  {a^{\mu} \partial_{\mu}  \over a}= \gamma v_{p} e^{-\lambda} {\partial \over \partial t} + \gamma {v_{c} \over v_{p}} {\partial \over \partial r}
\,. \label{dsigma}
\end{eqnarray}
Note that in the rest frame of the $i$:th fluid component the derivative with respect to $\tau$ reduces to $(- g_{tt})^{-1/2} \, \partial / \partial t$ and the derivative with respect to $\sigma$ to $g_{rr}^{-1/2}\,  \partial / \partial r  $.
Eq.~(\ref{cons2}) is actually a contracted form of the standard Euler equation, which reads $h^{\mu\nu} \partial_{\nu}p = -a^{\mu}  \,  (\rho+p)$.
From Eq.~(\ref{cons2}) it is clear that the four-velocity of a dust ($p=0$) component is geodesic.

By identifying the reference-frame velocity $u_{\rm rf}^{\alpha}$ with the  $\bar i$:th fluid-component velocity $u_{\bar i}^{\alpha}$ we can obtain the rest-frame expressions of Eqs.~(\ref{cons1}-\ref{cons2}) by simply setting $v_c=0$:
\begin{eqnarray}
 e^{-\lambda}  \dot{\rho}_{\bar i} & = & -  \Theta_{\rm rf} \, (\rho_{\bar i}+p_{\bar i})   \,, \\
 p_{\bar i} ' &=& - a_{r,\, \rm rf} \, (\rho_{\bar i}+p_{\bar i})   \,, \label{cons2e0}
\end{eqnarray}
respectively, where $a_{r,\, \rm rf}= \lambda'$ and $a_{t,\, \rm rf}=0$ (see Appendix \ref{exten}).
The last equation shows that the pressure gradients are the cause for the non-geodesity of the reference frame relative to the $\bar i$:th fluid component

Finally, because of the (twice contracted) Bianchi identity it holds $\nabla_{\mu}  G^{\mu\nu} =0$ and so:
\begin{equation}
\sum_{i=1}^{n} \nabla_{\mu} T^{\mu\nu}_{i}=0 \,.
\end{equation}
Consequently, two conservation equations will not be independent and can be discarded. Equivalently, we may replace, if convenient, any other two dynamical equations with the latter two extra conservation equations.

\subsection{Gravitating mass}

The effective gravitating total mass $F$ is given by
\begin{equation} \label{effmass}
   2 G  F \equiv H_A^2 Y^{3} - 2 E \, Y  \,,
\end{equation}
where $G$ is the gravitational constant.
This quantity corresponds to the mass in the Schwarzschild's metric if the metric of Eq.~(\ref{metric}) was matched to the latter at a given radius $r$. See Refs.~\cite{Cahill:1970a, Cahill:1970b, Alfedeel:2009ef} for more details.
It is interesting to note that Eq.~(\ref{effmass}) may be rewritten as:
\begin{equation} \label{energy}
E = {1 \over 2} \left( {d Y \over d\tau_{\rm rf}} \right)^{2} - {G F \over Y} \,,
\end{equation}
from which follows that the curvature function $E$ can be interpreted as the total energy per unit of mass of a shell, and that to it contribute the kinetic energy per unit of mass and the potential energy per unit of mass due to the total gravitating mass up to that shell.
Note that thanks to spherical symmetry one is able to define a potential energy also in cases far away from nearly Newtonian ones and that the potential energy is related to the curvature~\cite{Bondi:1947av}.

In giving the initial conditions it will be useful to divide the total mass $F$ into the separate masses of the $i$:th fluid components:
\begin{equation} \label{Fsplit}
F= \sum_{i=1}^{n} F_{i}  \,.
\end{equation}
Note, however, that this division is irrelevant for the dynamics as only the total $F$ enters the evolution equations.
By substituting Eqs.~(\ref{BTtt}-\ref{BTrr}) and (\ref{effmass}) into the (combinations of the) Einstein's Eqs.~(\ref{ee1}-\ref{ee2}) one obtains the equations for $\dot{F}$ and $F'$, which using Eq.~(\ref{Fsplit}) read:
\begin{eqnarray}
e^{-\lambda} \sum_{i=1}^{n} \dot{F}_{i} &=& -4 \pi Y^{3}  \sum_{i=1}^{n}  \rho_{i} \, \gamma_{i}^2 \Big [  H_{A} (v_{i,\,p}^{2}+w_{i}) +S_{A} ( 1+w_{i})  \Big]  \,,  \label{BAMd} \\
\sum_{i=1}^{n} F_{i}' &=& 4 \pi Y^{3} \sum_{i=1}^{n}   \rho_{i} \, {\gamma_{i}^2 \over v_{i,\, c}}  \Big [   H_{A} (1+ w_{i}) v^{2}_{i,\,p}+S_{A} (1+ v_{i,\,p}^{2} w_{i})  \Big]   \,,  \label{BAMp}
\end{eqnarray} 
where the angular ``spatial'' expansion rate $S_{A}=v_{c} \, {Y' \over Y}$ has been introduced in Appendix \ref{exten}.
Note that the Einstein's equations give Eqs.~(\ref{BAMd}-\ref{BAMp}); therefore to write Eq.~(\ref{Fsplit}) we have put the constants of integration to zero.
Note also that the evolution of the $i$:th mass $F_{i}$ is sourced by the gravitational influence of pressures and velocities of the other fluid components, differently from the case of the conservation equations (\ref{cons1}-\ref{cons2}) which apply to the fluid components individually: decoupled fluids are indeed still gravitationally coupled.
In the case of only one component Eqs.~(\ref{BAMd}-\ref{BAMp}) have a simple interpretation as discussed in Appendix \ref{pLTB}.

\subsection{Full set of equations}

The total number of independent equations is generally $4 +2 n -2$, where the first term is the number of nontrivial Einstein's equations, the second comes from the conservation equations and the last from the Bianchi identity. By including the definition of mass in Eq.~(\ref{effmass}) we have a total of $3 +2 n$ equations.
We choose to use all the $2n$ conservation equations, two of which will replace the acceleration equation (\ref{acceq}) and the evolution equation (\ref{BAMd}) for the total $F$.
We now list for clarity the full set of $3+2n$ exact equations governing the dynamics of a spherically symmetric inhomogeneous model with a multi-component ideal fluid:
\begin{eqnarray}
H_A^2 & = & \frac{2}{Y^3} \sum_{i=1}^{n}G F_{i}+ \frac{2 E}{Y^2}  \,,  \label{fset1} \\
{ e^{-  \lambda} \dot{E}  \over 1+2E}  &=&  {\kappa \over 2} {Y \over \sqrt{1+2E}} \sum_{i=1}^{n} \rho_i (1+w_i) v_{i,\,p} \gamma_i^2 
+  {e^{-  \lambda} \dot Y \over \sqrt{1+2E}} \, a_{\rm rf}   \,,    \label{fset2}\\
\sum_{i=1}^{n} F_{i}' &=& 4 \pi Y^{3} \sum_{i=1}^{n}    \rho_{i} \, {\gamma_{i}^2 \over v_{i,\, c}}  \Big [   H_{A} (1+ w_{i}) v^{2}_{i,\,p}+S_{A} (1+ v_{i,\,p}^{2} w_{i})  \Big]   \,,   \label{fset3}\\
 {d \rho_{i} \over d \tau_{i}}& = & -\Theta_{i} \, (\rho_{i} +p_{i}) \, \phantom{oooooooooo[}   i=1,\dots, n  \,,    \label{fset4}\\
{d p_{i}  \over d \sigma_{i}} &=& -  a_{i} \,  (\rho_{i} +p_{i}) \,  \phantom{oooooooooop} i=1,\dots, n   \,,   \label{fset5}
\end{eqnarray}
where the unknown functions are $Y$, $E$, the total $F$ and $n$ copies of $\rho_i$ and $v_{i,\,c}$.
The reference-frame velocity $u_{\rm rf}^{\alpha}$ can be taken to be geodesic so that $a_{\rm rf}=\lambda=0$.
Alternatively, one can identify the reference-frame velocity with the velocity of the $\bar i$:th fluid component, $u_{\rm rf}^{\alpha}=u_{\bar i}^{\alpha}$, with the consequence that $v_{\bar i,\,c}=0$ and $\lambda$ becomes an unknown function in its stead.
The equations of state $w_{i}$ have to be given as an external input.
From Eq.~(\ref{fset1}) it seems clear that the dynamics is similar to the FLRW one, the complication being an array of interconnected  equations defining the evolution of gravitating mass and curvature.

From Eq.~(\ref{fset1}) one may obtain the age of the universe at a given coordinate radius $r$. For example, in the case of expansion (positive root, $\dot Y >0$) it is:
\begin{equation} \label{aou}
t_{A}(r, t) \equiv t- t_{B}(r) = \int_{ t_{B}(r) }^{t} dt \; e^{-\lambda} \dot Y \left [ {2 G F(r, t) \over Y} +2E(r, t) \right ]^{-1/2} \,,
\end{equation}
which can be computed once the dynamical equations (\ref{fset1}-\ref{fset5}) are solved.
The quantity $t_{B}$ is the so-called bang function also present in LTB models.

\subsection{Initial and boundary conditions} \label{incond}

\paragraph{Initial conditions.}

We will give initial conditions at the time $\bar t < t_{0}$, where $t_{0}$ is the present time.
First we fix the residual gauge freedom in the coordinate $r$ by setting $Y(r, \bar t)= a(\bar t) r$, where $a(t)$ is the scale factor of the external FLRW model to which we will match the inhomogeneous metric (see below). If one is not interested in embedding the metric, $a(\bar t)$ may be taken as an arbitrary number.
Then, $E(r, \bar t)$, $F(r, \bar t)$ and $n$ copies of $\rho_i(r, \bar t)$ and $v_{i,\,c}(r, \bar t)$ are needed.

The curvature function $E$ may be specified by demanding a homogeneous age of the universe $t_{A}$ at $\bar t$, which gives an additional constraint between the total $F$ and $E$.
From Eq.~(\ref{aou}), in the approximation that $F(r,t)\simeq F(r,\bar t)$ and $E(r,t) \simeq E(r,\bar t)$ for~$t< \bar t$, we can obtain:
\begin{equation} \label{hbb}
t_{A}(r, \bar t) \equiv \bar t- t_{B}(r) \simeq \int_{0}^{Y(r, \bar t)} dy \; e^{-\lambda(r,\bar t)}\left [ {2 G F(r,\bar t) \over y} +2E(r,\bar t) \right ]^{-1/2} \,,
\end{equation}
which can be used without solving the dynamical equations to demand at $\bar t$ the homogeneous big bang, $t_{B}(r)=0$.

Note then that in the set of Eqs.~(\ref{fset1}-\ref{fset5}) there is not a dynamical equation for $F$, which can always be found by integrating Eq.~(\ref{fset3}) for a constant $t$ once the boundary condition $F(0,t)=0$ is given.
In particular, Eq.~(\ref{fset3}) relates the initial conditions for $F$, $\rho_i$ and $v_{i,\,c}$, which cannot be given independently, as it is intuitively clear.
It is important to point out that while the initial conditions can be given individually for the different fluids by considering Eq.~(\ref{fset3}) as $n$ independent equations, this will not be true at later times for which only the total $F$ matters.
In other words, one cannot obtain the individual $F_{i}$ by integrating the corresponding $i$-piece of Eq.~(\ref{fset3}) and then combining them to obtain $F$.
This also means that for $t > \bar t$ there is not a meaningful way to define the individual $F_{i}$, which are not physically relevant.
We can instead define the invariant mass $M$ as shown, in the example of Section \ref{colla}, by Eq.~(\ref{mmass}).

If we identified the reference-frame velocity with the velocity of the $\bar i$:th fluid component, $u_{\rm rf}^{\alpha}=u_{\bar i}^{\alpha}$, we may have to give initial conditions for $\lambda$ (and not $v_{\bar i,\,c}$ which is identically zero).
Note, however, that $\dot \lambda$ only appears in the acceleration equation which we have discarded.
As with $F$, it is therefore enough to give a boundary condition (see below) in order to obtain with Eq.~(\ref{cons2e0}) the lapse function at $\bar t$ for any $r$.

We have now all the functions and their $r$-derivatives on the hypersurface of constant $\bar t$. 
The next step is to evolve the model forwards in time along the worldlines of constant $r$, which is done by simultaneously integrating the first-order in $t$ differential equations listed in (\ref{fset1}-\ref{fset5}).
While the $t$-derivatives for $Y$, $E$ and $\rho_i$ are evident, $\dot v_{i,\, c}$ enters only through the expansion and acceleration scalars.
Finally, $\lambda(r,t)$ and $F(r,t)$ are found by integration for the needed constant $t$.
The equations that we have discarded thanks to the Bianchi identity may be used to cross-check the accuracy of the numerical integration.

\paragraph{Boundary conditions.}

In discussing the necessary boundary conditions we will particularize the analysis to a metric which is matched to an external FLRW model at some comoving coordinate radius $r_{b}$.
We will give boundary conditions at the initial time: the evolution equations will automatically maintain them at later times.
Similarly to the embedding of an LTB model, the curvature and the integrated gravitating masses have to match the FLRW values, $E(r_{b}, \bar t) =- {1 \over 2} k \, r_{b}^{2}$ and $F_{i}(r_{b},\bar t)=  F_{i}^{\rm out}(r_{b}, \bar t)$, respectively (in the chosen gauge the scale function is already matched as $Y(r, \bar t)= a(\bar t) r$).
The latter condition for the case of $v_{i,\,c}(r, \bar t)=0$ explicitly demands that $\rho_{i}^{\rm out} (\bar t)  = {3 \over r_{b}^{3}}   \int_{0}^{r_{b}}  \rho_{i}(\hat r, \bar t) \,{\hat r}^{2}  d \hat r$, as it is easy to see from Eq.~(\ref{fset3}).
According to the Darmois-Israel boundary conditions \cite{HellabyJC} the local density may be discontinuous through a timelike surface (constant $r_{b}$), but the pressure must be continuos  $p_{i}(r_{b},\bar t)=  p_{i}^{\rm out}(\bar t)$.
Furthermore the  peculiar velocities have to vanish, $v_{i,\,c}(r_{b}, \bar t)=0$, as also the lapse function, $\lambda(r_{b}, \bar t)=0$, as we want to describe the FLRW model in a geodesic reference frame where proper and coordinate times coincide.
Even if not necessary it may be desirable to match smoothly the background, i.e., $\rho_{i}(r_{b},\bar t)=  \rho_{i}^{\rm out}(\bar t)$ and $\rho_i'(r_{b}, \bar t)=p_i'(r_{b}, \bar t)=\lambda'(r_{b}, \bar t)=v_{i,\,c}'(r_{b}, \bar t)=0$.

Finally, we discuss the regularity conditions at the center of the inhomogeneous metric.
Because we do not want a singularity at the center it is $F_{i}(0, \bar t)=E(0, \bar t)=0$. In particular, near the origin it is $F_{i} \propto r^{3}$ and $E \propto r^{2}$, as one can see from Eq.~(\ref{effmass}) where $Y\propto r$ and a finite $H_{A} \neq 0$ is assumed (see also \cite{Mustapha:1998jb}).
The latter also implies $\dot Y \propto r$ and so $\dot Y(0, \bar t) = Y(0, \bar t) =0$.
Also vanishing at the center must be the velocities, $v_{i,\,c}(0, \bar t)=0$, and the pressure gradient, $p_i'(0, \bar t)=0$, which implies $\lambda'(0, \bar t)=0$.
It seems again desirable to have densities and velocities smooth at the origin, $\rho_i'(0, \bar t)=v_{i,\,c}'(0, \bar t)=0$.

\subsection{Light propagation}

The metric (\ref{metric}) implies that for radial null worldlines it is
\begin{equation}
   \frac{dt}{du} = \pm \frac{dr}{du} e^{-\lambda} \frac{Y'}{\sqrt{1+2E}}\,, \label{lc}
\end{equation}
where $u$ is an affine parameter. Let us consider a photon emitted at $t(u)$ in a coordinate time $\nu(u)$ (emitting source and observer are taken to be in the rest frame).
Expanding the right-hand side of Eq.~(\ref{lc}) around $t(u)$ and keeping terms up to first order in $\nu(u)$ we get
\begin{equation}
   \frac{dt}{du} + \frac{d\nu}{du} = -\frac{dr}{du} e^{-\lambda} \frac{Y'}{\sqrt{1+2E}} - \frac{dr}{du}  \frac{Y'}{\sqrt{1+2E}} \left(  H_{R} - e^{-\lambda}  \dot{\lambda} \right) \nu \,,
\end{equation}
which, using Eq.~(\ref{lc}) for $dt/du$, gives
\begin{equation}
    \frac{1}{\nu}\frac{d\nu}{du} = - \frac{dr}{du}  \frac{Y'}{\sqrt{1+2E}} \left( H_R - e^{-\lambda} \dot{\lambda} \right) \,.
\end{equation}                                     
The redshift is defined as\footnote{This definition is different from the one in \cite{Lasky:2010vn}. The authors of \cite{Lasky:2010vn} agree, however, that (\ref{zeq}) is correct.}
\begin{equation} \label{zeq}
   z = \frac{e^{\lambda_0}\nu_0 - e^{\lambda}\nu}{e^{\lambda}\nu}\,,
\end{equation}
where the subscript $0$ refers to values evaluated at the affine-parameter value $u_0$ at the observer's position. The derivative of $z$ with respect to $u$ is then
\begin{equation}
   \frac{dz}{du} = -(1+z) \frac{dr}{du}\left( \frac{d\lambda}{du} + \frac{1}{\nu}\frac{d\nu}{du} \right) 
                 = -(1+z) \frac{dr}{du}\left( \lambda' - \frac{Y'}{\sqrt{1+2E}}H_R \right) .   \label{dz}
\end{equation}
Using (\ref{dz}) and remembering the expression for the rest-frame acceleration scalar (see Appendix \ref{exten}), Eq.~(\ref{lc}) gives a pair of equations  
\begin{equation} 
           \frac{dt}{dz} = \frac{1}{1+z}   \,   \frac{e^{-\lambda}}{ a_{\rm rf} - H_R} \,, \qquad
           \frac{dr}{dz} = -\frac{1}{1+z}    \,   \frac{1}{a_{\rm rf} - H_R} \frac{\sqrt{1+2E}}{Y'} \label{lc2}  \,,
\end{equation}
which determines the coordinates $t$ and $r$ on the light cone as functions of the redshift, and can be put in the following gauge invariant form:
\begin{equation} 
           \frac{d\tau_{\rm rf}}{dz}= \frac{1}{1+z}   \,   \frac{1}{ a_{\rm rf} - H_R} \,, \qquad
           \frac{dr_{p}}{dz} = -\frac{1}{1+z}    \,   \frac{1}{a_{\rm rf} - H_R}  \,,
\end{equation}
where $r_{p}$ is the proper radial distance.
In the $\Lambda$LTB model where $\dot E={a_{\rm rf}}=0$, the previous equations reduce to the familiar form
\begin{equation}
            \frac{dt}{dz} = -\frac{Y'}{(1+z)\dot{Y}'} \,, \qquad
            \frac{dr}{dz} =\frac{\sqrt{1+2E}}{(1+z)\dot{Y}'}   \,.
\end{equation}

\section{Two non-comoving dust components in a flat \boldmath $\Lambda$CDM universe} \label{colla}

As a first application of the formalism introduced in this paper, we will consider an inhomogeneous sphere embedded into a flat $\Lambda$CDM universe.
The inhomogeneities will be given by two non-comoving dust components, one of which may be interpreted as baryonic matter and the other as dark matter. The reader should keep in mind, however, that these results are merely illustrative of the multi-fluid model here presented.
This solution becomes the usual $\Lambda$LTB model if the density of the second dust component is set to zero or if their relative velocity is set to zero.

We will use the reference frame comoving with the component labelled by ``$M$'', so that the relative Euler equation gives $\lambda=0$.
The other conservation equation reads:
\begin{equation} \label{consam}
{\dot \rho_{M} \over \rho_{M}}   = - \left ( H_R  +  2 H_A \right)  =  - { \left( J_{R} J_{A} \right)\dot{}  \over J_{R} J_{A} } \,,
\end{equation}
and can be directly integrated to give:
\begin{equation} \label{dustevo}
{\rho_{M}(r, t) \over \rho_{M}(r, \bar t)}   =   {J_{R}(r, \bar t) J_{A}(r, \bar t)  \over  J_{R}(r, t) J_{A}(r, t)}     =  {Y^{2}(r, \bar t) Y'(r, \bar t)  \over \sqrt{1+2E(r,\bar t)}}      { \sqrt{1+2E(r, t)}   \over   Y^{2}(r, t) Y'(r, t) } \,,
\end{equation}
where $\bar t$ is the initial time.
We will label the other dust component with ``$N$'' and use the simpler notation $\gamma_{N}=\gamma$, $v_{N,\, p}=v_{p}$, $v_{N,\, c}=v_{c}$ as the corresponding $M$ quantities are trivial.
About $\Lambda$ note that the corresponding conservation equations (\ref{fset4}-\ref{fset5}) trivially state that $\rho_{\Lambda}$ is uniform and constant.

\subsection{Initial and boundary conditions}

First we set the flat $\Lambda$CDM model by fixing the present-day expansion rate to $H_{0}=100  h$ km s$^{-1}$ Mpc$^{-1}$ with $h=0.7$ and the matter density parameter to $\Omega_{\rm matter}=0.3$.
We then solve the background equations to find the evolution of the scale factor $a(t)$ and so the initial time $\bar t$ corresponding to $\bar z={a(t_{0}) \over a(\bar t)}-1$, which we will fix illustratively to $\bar z=5$.
By redshift $z$ we will always mean the redshift relative to an observer in the background model, which will differ, for example, from the redshift relative to the observer at the center of the spherical inhomogeneity.
The inhomogeneous sphere will have a comoving radius of $r_{b}=100 h^{-1}$~Mpc.

Next we give initial and boundary conditions at $\bar t$.
For the scale function we set $Y(r, \bar t)= a(\bar t) r$.
Then we have to give initial conditions for $F(r, \bar t)=\bar F_{M}(r)+\bar F_{N}(r)+\bar F_{\Lambda}(r)$.
For $\Lambda$ it is simply $\bar F_{\Lambda}(r)= {4 \pi \over 3}  a^{3}(\bar t) r^{3} \rho_{\Lambda}$.
We stress again that the division of the total $F$ into separate components is meaningful only at the initial time as only the total gravitating mass $F$ enters the evolution equations.
Note, however, that it is possible to define {\em individual} invariant mass components $M_{i}$. For instance, for the $M$ component it is:
\begin{equation} \label{mmass}
M_{M}' = {4 \pi Y^{2} Y' \over \sqrt{1+2E}}   \; \rho_{M} \,,
\end{equation}
from which it is easy to verify using Eq.~(\ref{consam}) that it is correctly $\dot M_{M}=0$ (while $\dot F \neq 0$).

The remaining background matter density is split between the $M$ and $N$ components according to a fraction $q$ such that it is $\rho_{M}^{\rm out} (\bar t) = q \, \rho_{\rm matter}^{\rm out} (\bar t)$ and $\rho_{N}^{\rm out} (\bar t) = (1-q) \, \rho_{\rm matter}^{\rm out} (\bar t)$.
We choose the profile of the $M$ component to be:
\begin{equation} \label{rhoMb} 
{\rho_{M} (r, \bar t)  \over \rho_{M}^{\rm out} (\bar t)} =1+ \delta_{1}(r,r_{t_{M}},\delta_{M1,M2},\alpha_{M1,M2})
\end{equation}
where the $\delta_{n}$ function is given in Appendix \ref{wdn} and $\delta_{M}=\delta_{M1}+\delta_{M2}$ is the contrast in matter density between the center of the spherical inhomogeneity and the background value of $\rho_{M}^{\rm out}$.
This inhomogeneous profile satisfies the smoothness conditions $\rho_{M}'(0, \bar t)=\rho_{M}'(r_{b}, \bar t)=0$ and continuously matches the background density $\rho_{M} (r_{b}, \bar t) = \rho_{M}^{\rm out} (\bar t)$.
Moreover, the matching requires that $\bar F_{M}(r)$ has the corresponding background value and so the following condition has to be satisfied:
\begin{equation} \label{matchM}
 \rho_{M}^{\rm out} (\bar t)  = {3 \over r_{b}^{3}}   \int_{0}^{r_{b}}  \rho_{M}(\hat r, \bar t) \,{\hat r}^{2}  d \hat r   \,,
\end{equation}
from which follows that it has to be $\delta_{M1} \, \delta_{M2}<0$.
$\delta_{M1}>0$ will give an overdensity, while $\delta_{M1}<0$ an underdensity.
For a given model we will fix the quantities $\delta_{M1,\,M2}$ and $\alpha_{M1,\,M2}$ and we will get analytically the radius $r_{t_{M}}$ by demanding Eq.~(\ref{matchM}).
The gravitating mass is then simply $\bar F_{M}(r)= 4 \pi a^{3}(\bar t)   \int_{0}^{r}  \rho_{M} (\hat r, \bar t) \hat r^{2} d \hat r$, which has also an analytic expression.

Next we need $\bar F_{N}(r)$, which we will take to be:
\begin{equation}
\bar F_{N}(r) = \bar F_{N}^{\rm out}(r) + \delta_{3}(r,r_{t_{N}},\delta_{N1,N2},\alpha_{N1,N2}) \,,
\end{equation}
with $\delta_{N2}=-\delta_{N1}$ so that $\bar F_{N}(r)$ is correctly matched to the background value of $\bar F_{N}^{\rm out}(r)= {4 \pi \over 3}  a^{3}(\bar t) r^{3} \rho_{N}^{\rm out} (\bar t) $.

We can now find the initial curvature function $E$ by demanding that the universe has the same age $\bar t$ for any $r$, i.e., from Eq.~(\ref{hbb}):
$t_{A}(r,\bar t) = \bar t $.
The approximation from neglecting the time dependence of $F$ and $E$ for $t<\bar t$ is very good as the cosmological constant is negligible at early times.

Next, we have to set the peculiar velocities for the $N$ component with respect to the $M$ component. As there is not pressure, the case $v_{c}(r,\bar t)=0$ gives trivially the dynamics of the $\Lambda$LTB model with a density profile that is the average of the $M$ and $N$ density profiles.
We choose the initial velocity profile to be:
\begin{equation}
v_{c}(r,\bar t)= \delta_{1}(r,r_{t_{v}},\delta_{v1,v2},\alpha_{v1,v2})
\end{equation}
with $\delta_{v2}=-\delta_{v1}$ so that there are no peculiar velocities at the center and at the border. The contrast $\delta_{v2}$ will give the maximum peculiar velocity in $c$ units.
Finally we can compute the initial density profile for the $N$ component, which from Eq.~(\ref{fset3}) is:
\begin{equation}
\rho_{N} (r, \bar t) = {\bar F_{N}'(r) \over 4 \pi \bar Y^{3}   \, {\bar \gamma^2 \over \bar v_{c}}  \Big (  \bar H_{A} \bar v^{2}_{p}+ \bar S_{A} \Big) } \,,
\end{equation}
where $\bar H_{A}$ is given by Eq.~(\ref{fset1}) evaluated at the initial time.
This terminates the necessary initial conditions.

\subsection{Evolution}\label{resu}

\begin{figure}
\begin{center}
\includegraphics[width= .49\textwidth,height=6.3 cm]{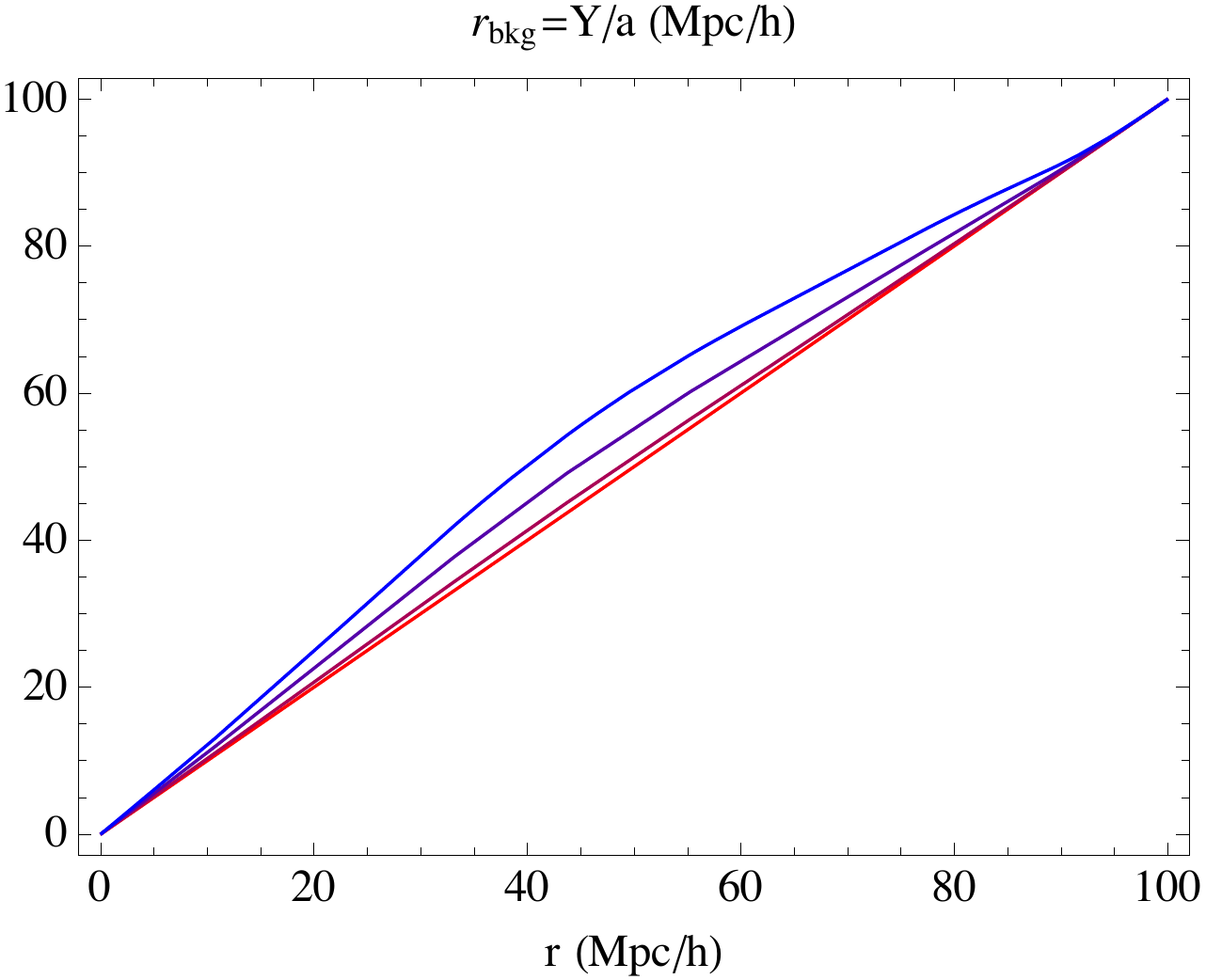}
\includegraphics[width= .49\textwidth]{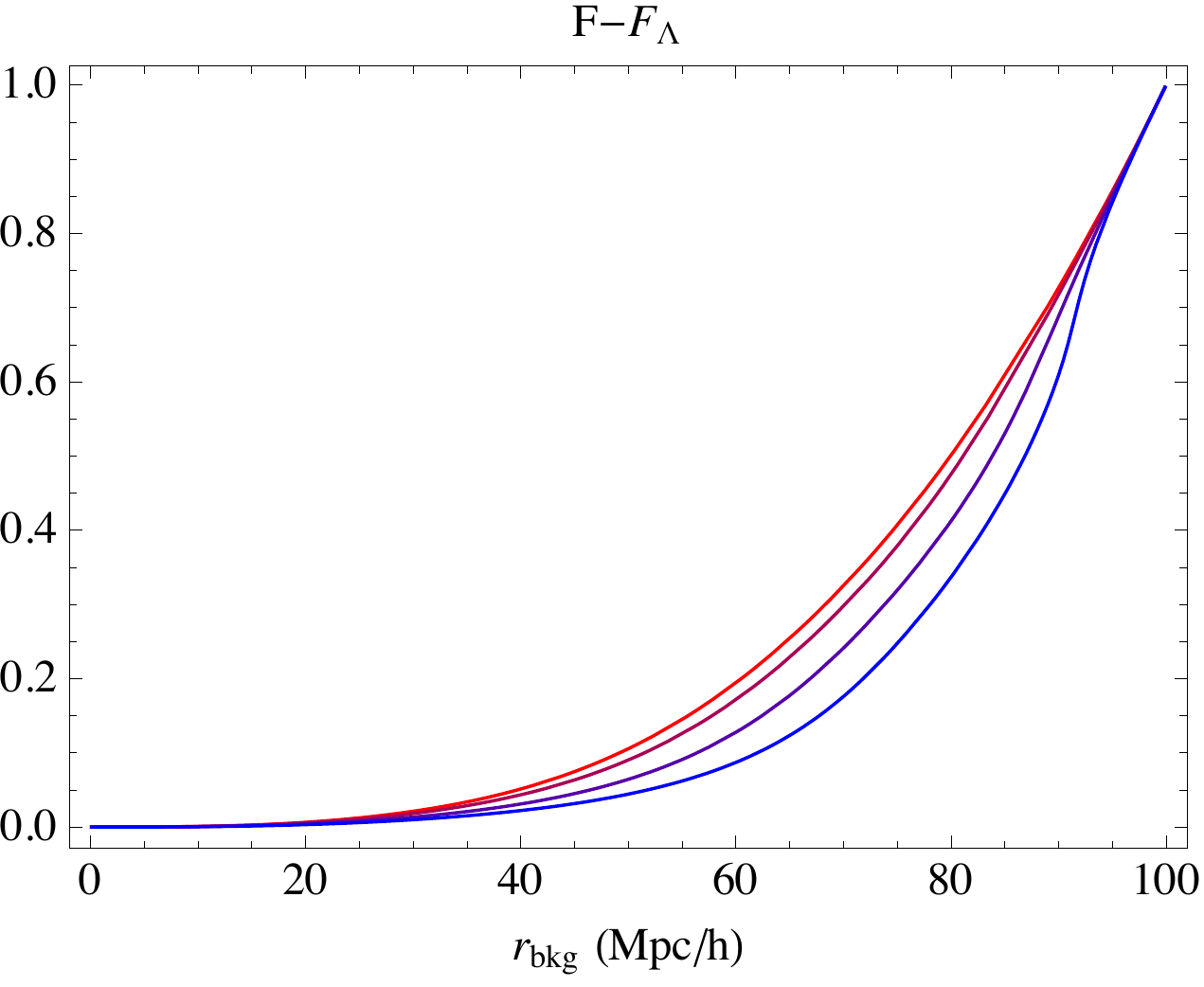}
\caption{
Left panel: evolution of the scale function $Y$ for the times corresponding to the background redshifts of $z=$5 (red), 3, 1 and 0 (blue). In particular, this plot shows the relation between the coordinate radius $r$ and the background comoving radius $r_{\rm bkg}=Y/a$ which will be used in all the other plots.
Right panel: Evolution of the gravitating mass $F$ for $z=$5 (red), 3, 1 and 0 (blue) with the contribution due to $\Lambda$ subtracted and normalized to unity at hole radius $r_{b}$. The time evolution is purely an effect of the peculiar velocities between the two dust components.
See Section \ref{resu} for more details.
}
\label{fig1}
\end{center}
\end{figure}

We will now show the exact general relativistic evolution of the system presented in the previous Section for the case of a central underdensity.\footnote{The actual parameters we use are $q=0.7$, $\alpha_{M1}=\alpha_{M2}=0.4$, $\delta_{M1}=-0.3$, $\delta_{M2}=0.05$,
$r_{t_{N}}=2 r_{b}/3$, $\alpha_{N1}=0.1$, $\alpha_{N2}=0$, $\delta_{N1}=-\delta_{N2}=0.1 \,  \bar F_{N}^{\rm out}(r_{b}/2)$,
$r_{t_{v}}=2 r_{b}/3$, $\alpha_{v1}=\alpha_{v2}=0$, $\delta_{v2}=-\delta_{v1}=0.5\cdot 10^{-2} /a(\bar t)$.}
The left panel of Fig.~\ref{fig1} shows the evolution of the scale function $Y$: in the interior region the scale function is expanding faster because of the less matter and consequently negative curvature present, as also shown by the spatial Ricci scalar ${\cal{R}}$ in the left panel of Fig.~\ref{fig2} (see Appendix \ref{pLTB} for its definition).
All the other figures will be plotted with respect to the background comoving radius defined as $r_{\rm bkg}=Y(r,t)/a(t)$.
This is a good definition as the curvature function is $E \ll 1$ (see right panel in Fig.~\ref{fig2}) and thus the proper radial distance is $r_{p}= \int  {Y' d\hat r\over \sqrt{1+2E}} \simeq Y$.

The evolution of the scale function $Y$ is governed by the gravitating mass $F$ and the curvature function $E$, as shown by Eq.~(\ref{fset1}).
The evolution of the gravitating mass $F$ is plotted in the right panel of Fig.~\ref{fig1}. We have subtracted the time-dependent component due to $\Lambda$, which is $F_{\Lambda}= {4\pi \over 3} Y^{3} \rho_{\Lambda}$, in order to show the genuine time dependence of the gravitating mass due to the dust components. As there is no pressure, the evolution displayed in the plot is due to the peculiar velocities between the two components.
In particular the gravitating mass at half the radius is decreasing in accordance with the outgoing flow shown in the right panel of Fig.~\ref{fig4}.

\begin{figure}
\begin{center}
\includegraphics[width= .49\textwidth]{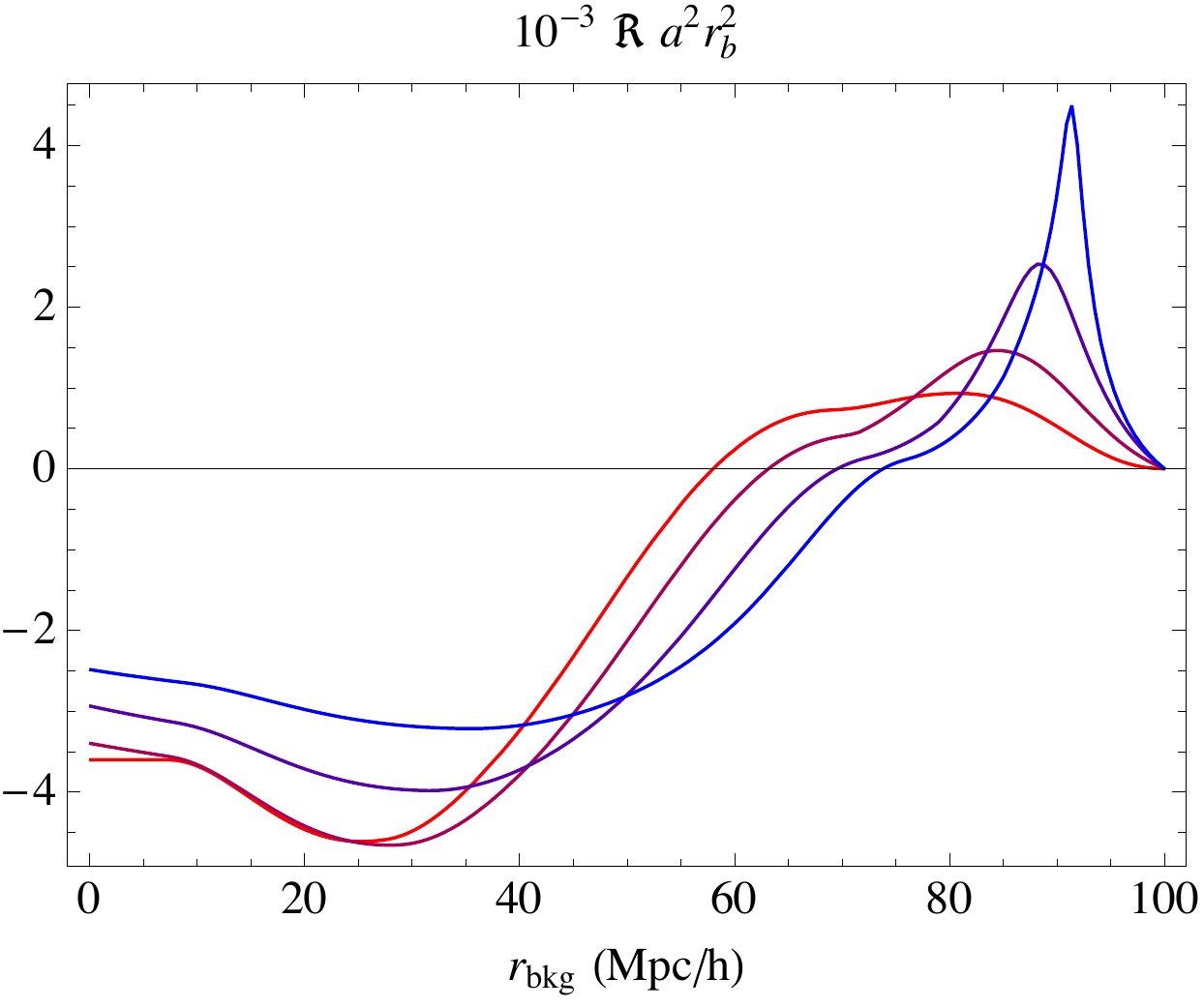}
\includegraphics[width= .49\textwidth,height=6.3 cm]{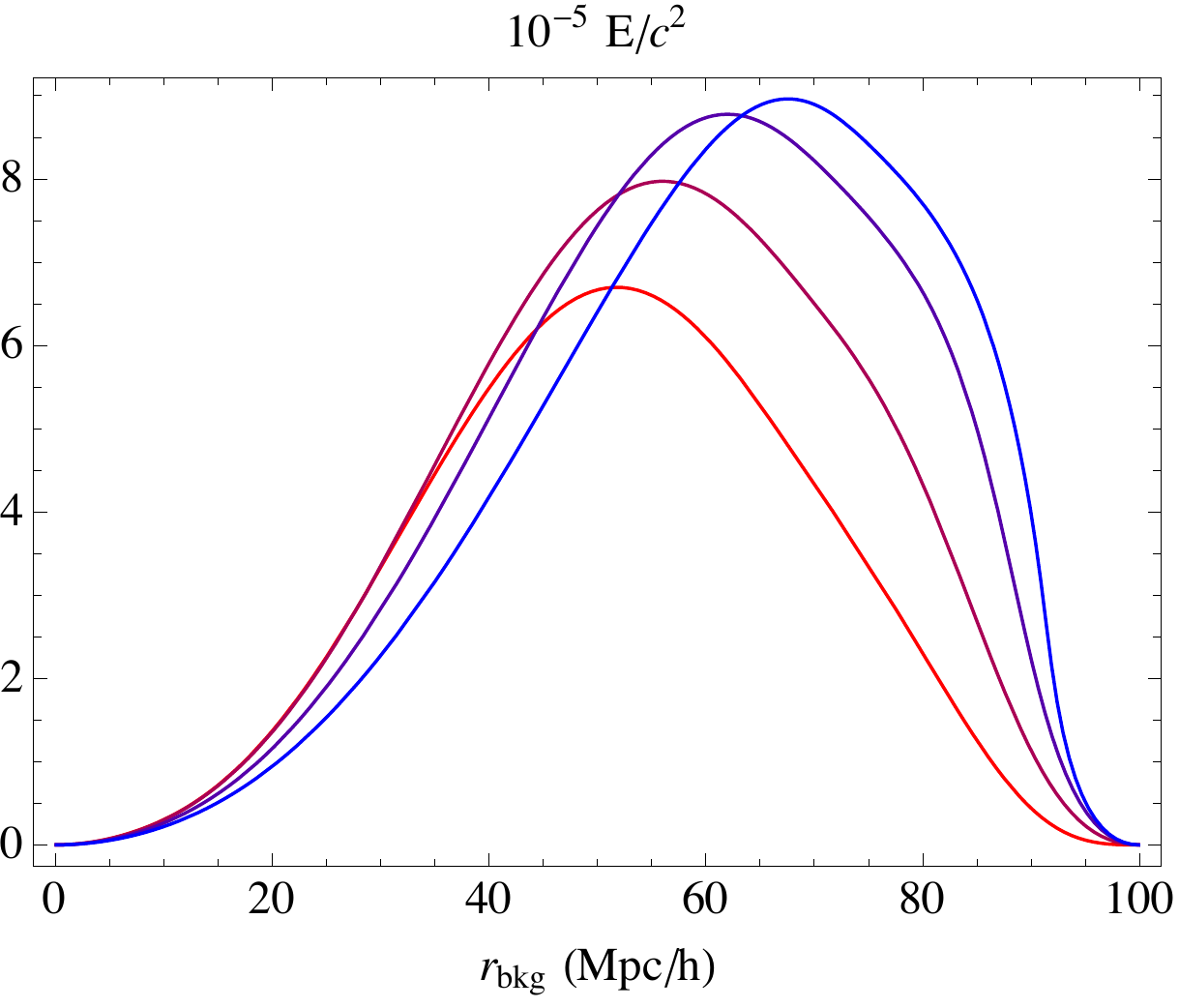}
\caption{
Left panel: evolution of the spatial Ricci scalar ${\cal{R}}$ plotted as the dimensionless combination ${\cal{R}} \, a^{2}r_{b}^{2}$ for $z=$5 (red), 3, 1 and 0 (blue). Its profile is closely related to the density profiles of the two dust components displayed in Fig.~\ref{fig3}.
Right panel: evolution of the curvature/energy function $E$ for $z=$5 (red), 3, 1 and 0 (blue). The plot shows an outgoing flow of energy in accordance with the peculiar velocity field shown in the right panel of Fig.~\ref{fig4}.
See Section \ref{resu} for more details.
}
\label{fig2}
\end{center}
\end{figure}

\begin{figure}
\begin{center}
\includegraphics[width= .49\textwidth]{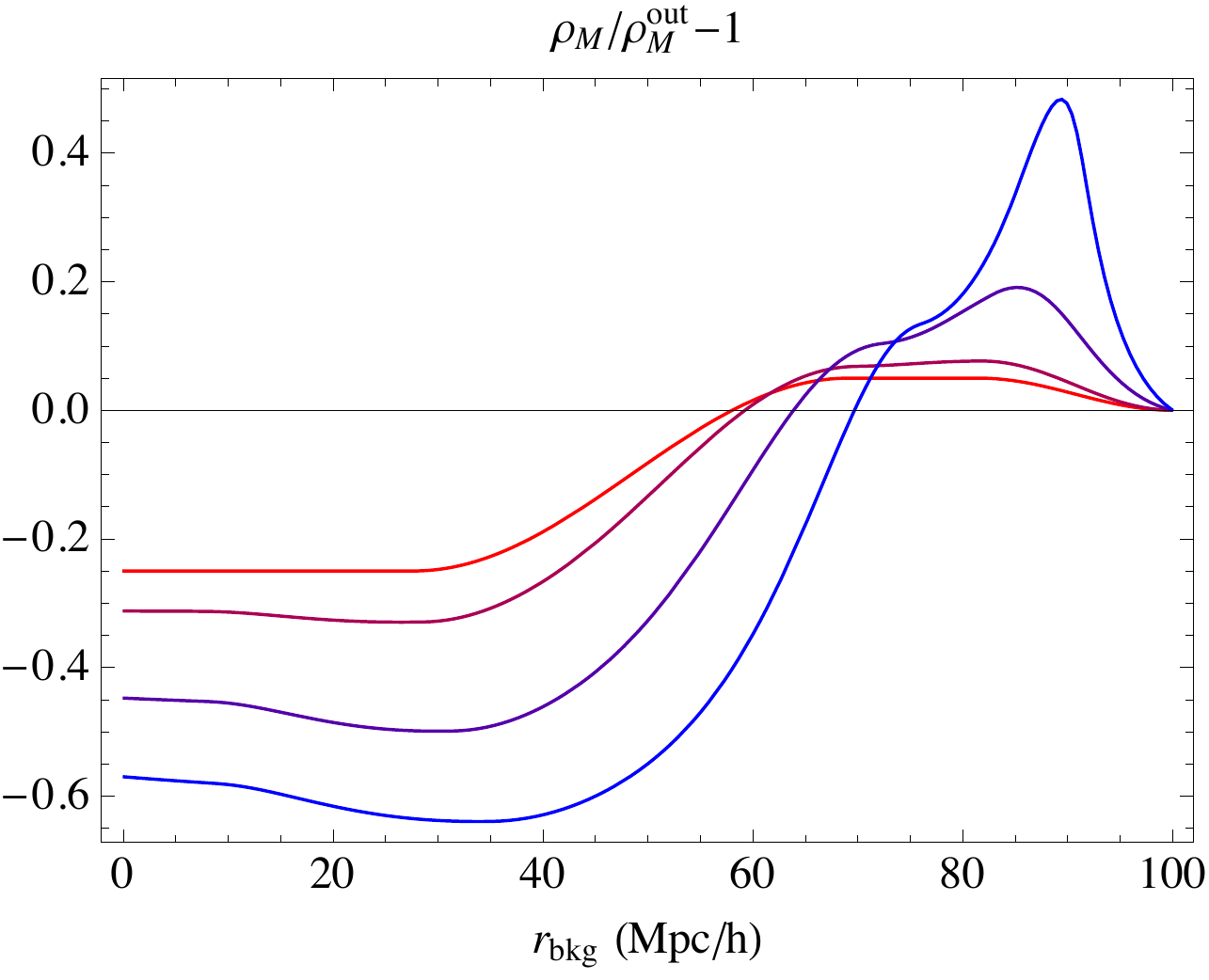}
\includegraphics[width= .49\textwidth, height=6.15 cm]{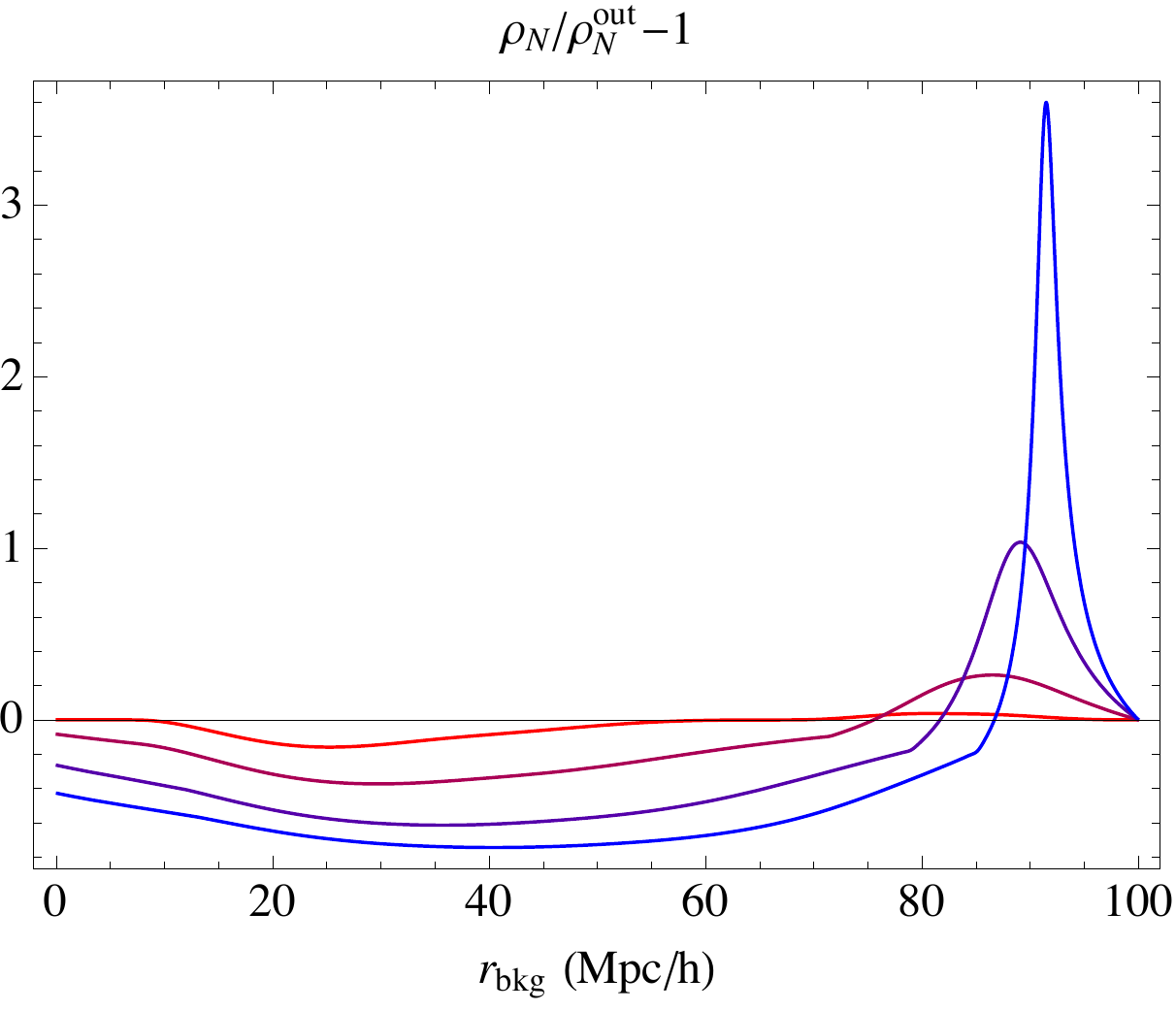}
\caption{
Evolution of the $M$ component local density (left panel) and $N$ component local density (right panel) for $z=$5 (red), 3, 1 and 0 (blue).
Both the density contrasts show a realistic evolution, with overdense regions contracting and becoming thin shells (mimicking structures), and underdense regions becoming larger and deeper (mimicking voids).
Note in particular how the $N$ component induces gravitationally a peak in the density of the $M$ component.
See Section \ref{resu} for more details.
}
\label{fig3}
\end{center}
\end{figure}

\begin{figure}
\begin{center}
\includegraphics[width= .49\textwidth,height=6.45 cm]{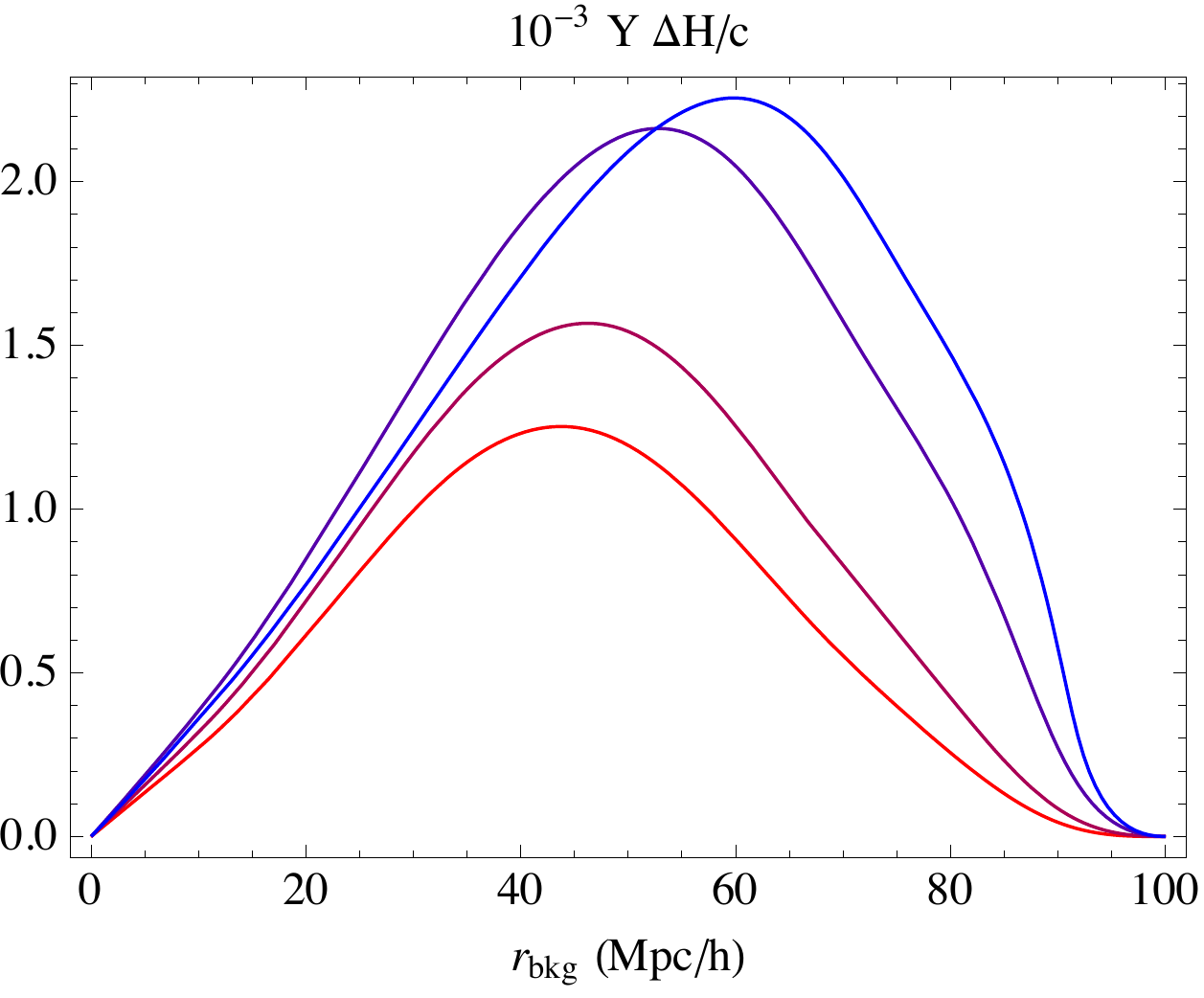}
\includegraphics[width= .49\textwidth]{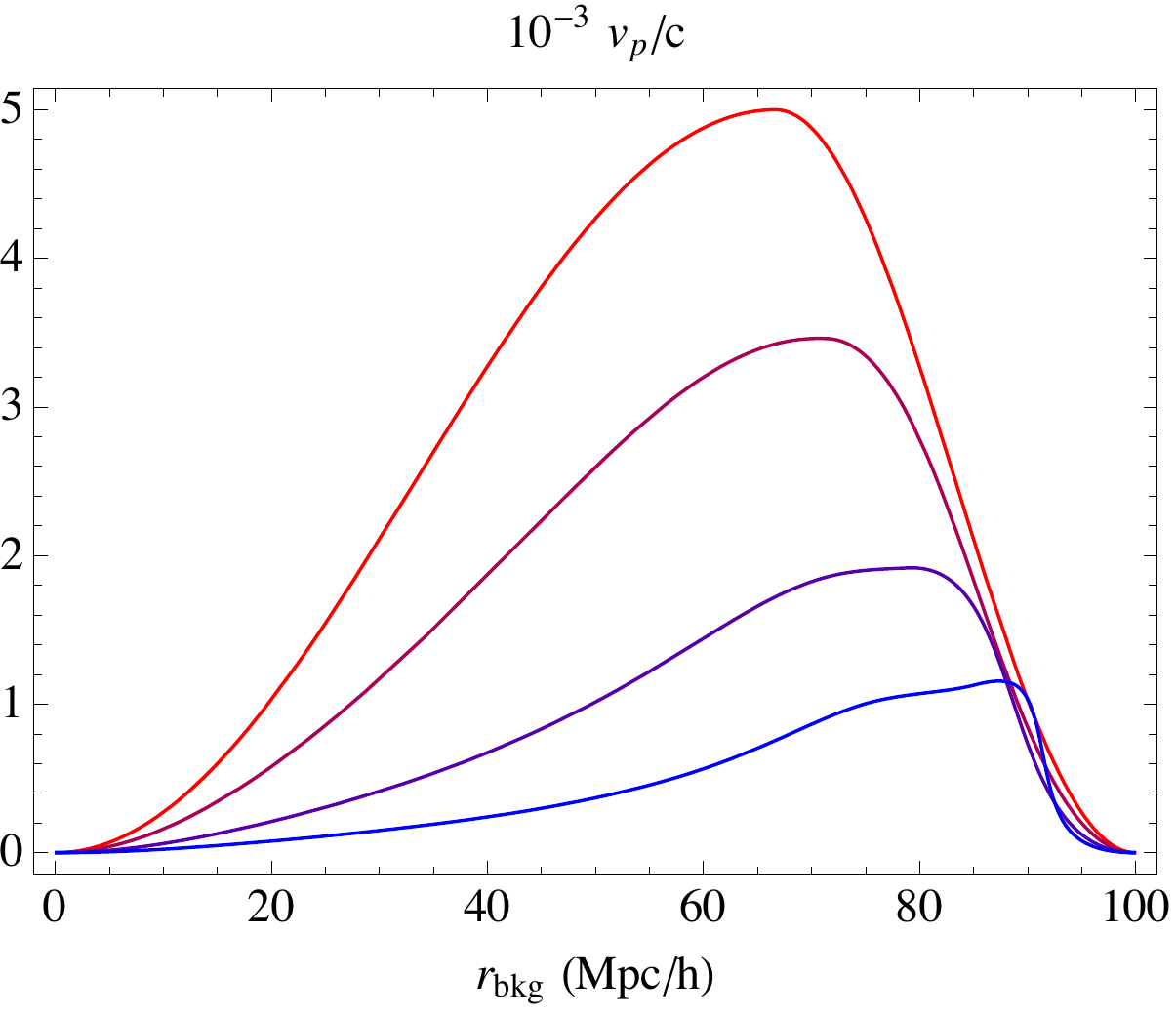}
\caption{
Left panel: evolution of the peculiar velocities $v_{M} \equiv Y (H_{A}-H_{\rm out})$ of the $M$ component with respect to the background for $z=$5 (red), 3, 1 and 0 (blue).
Velocities are increasing as the underdensity becomes emptier.
Right panel: evolution of the peculiar velocities of the $N$ component with respect to the $M$ component for $z=$5 (red), 3, 1 and 0 (blue).
The gravitational attraction between the two components reduces their relative velocities.
See Section \ref{resu} for more details.
}
\label{fig4}
\end{center}
\end{figure}

The evolution of the curvature function $E$, is plotted in the right panel of Fig.~\ref{fig2}. This is again a genuine effect of the peculiar velocities between the two dust components, as a cosmological constant never sources the evolution of the curvature function $E$. We remind that $E$ can be interpreted as the total energy per unit of mass of a given shell (see Eq.~(\ref{energy})), and one can see from the plot that  energy seems to flow outwards, again in agreement with the peculiar velocity field shown in the right panel of Fig.~\ref{fig4}. 
As said before, the left panel of Fig.~\ref{fig2} shows the Ricci scalar ${\cal{R}}$, which seems to follow a profile which is an average of the density contrasts of the two dust components, displayed in Fig.~\ref{fig3}.
We also remind that the curvature function is related to the Euclidean average of ${\cal{R}}$, as pointed out in Appendix \ref{pLTB}.

Both the density contrasts of the two dust components show a realistic evolution, as one can see from Fig.~\ref{fig3}.
Overdense regions start contracting and they become thin shells (mimicking structures), while underdense regions become larger and deeper (mimicking voids), and eventually they occupy most of the volume.
An interesting feature of this multi-component model is that the gravitational interaction between the two fluids is evident. The profile of the $M$ component (left panel) develops indeed a peak exactly where the $N$ component peaks: that is, an overdensity of one components drags in the collapse the other component.
This can be checked by decreasing the matter density allocated to the $N$ component, with the effect that the the peak at $90 h^{-1}$ Mpc in the $M$ component disappears.

\begin{figure}
\begin{center}
\includegraphics[width= .49\textwidth,height=6.4 cm]{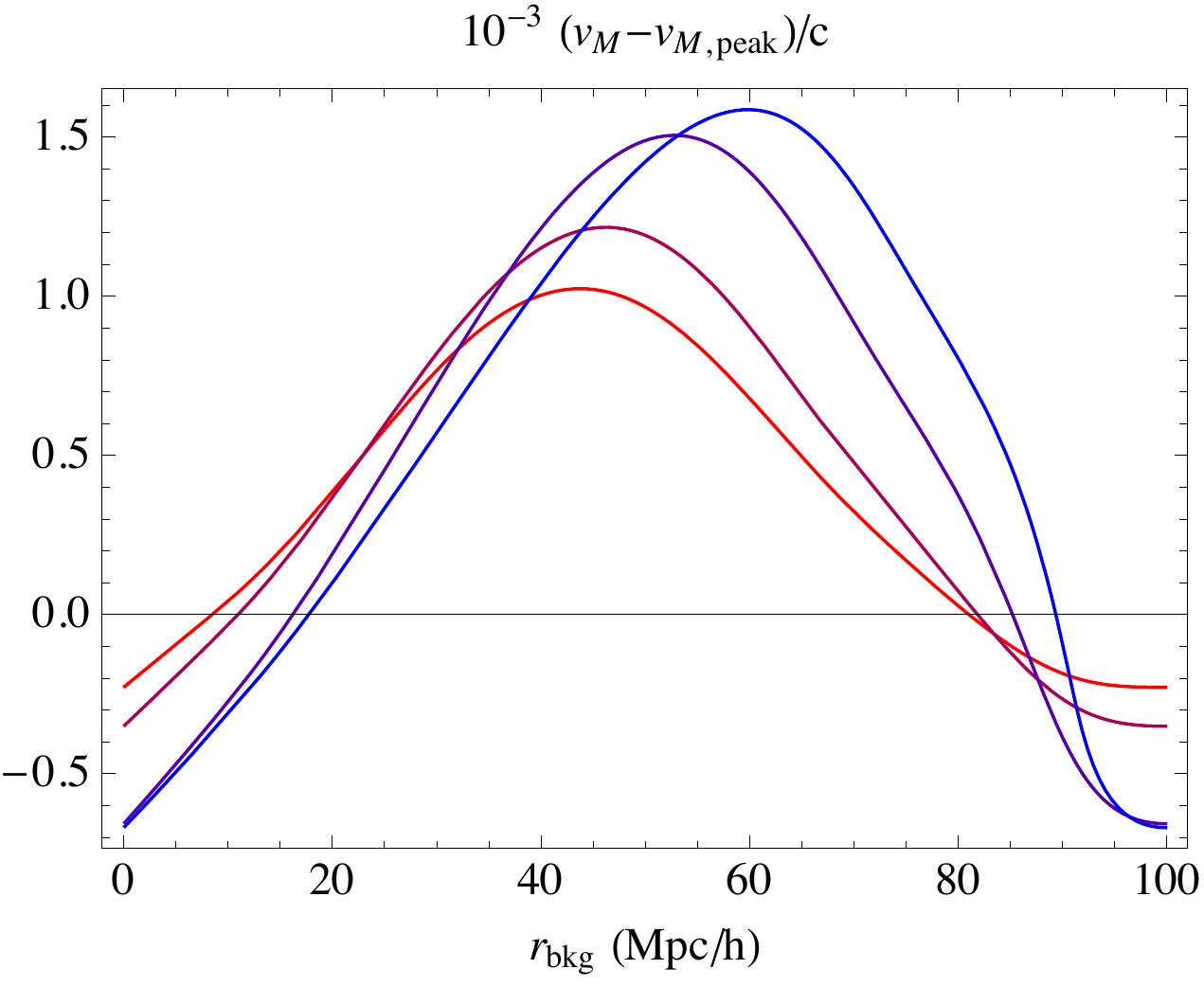}
\includegraphics[width= .49\textwidth]{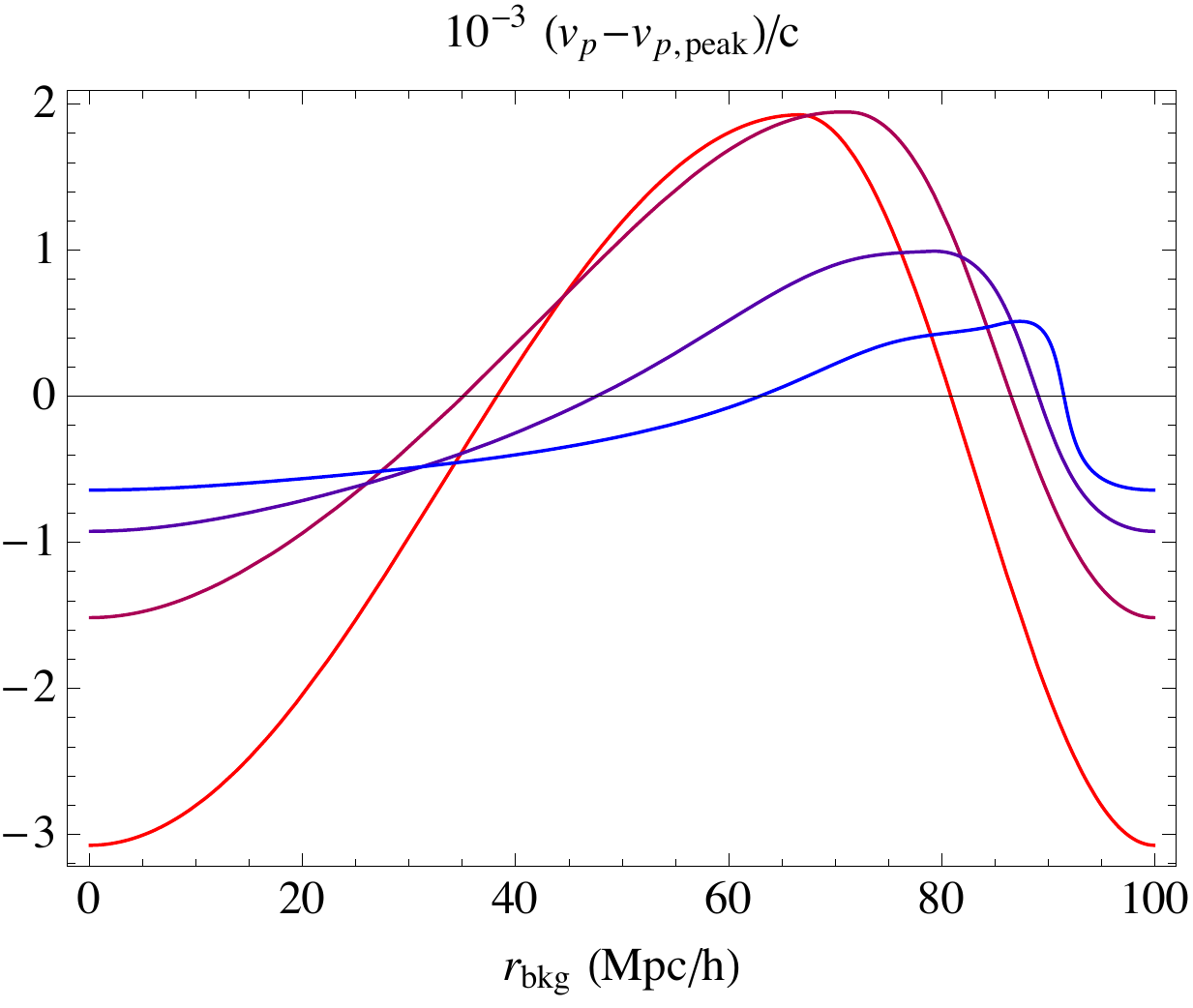}
\caption{
Same as Fig.~\ref{fig4} but with the velocity of the peak subtracted.
These plots show that the velocity flow near the peaks in density is natural: matter is falling towards the overdense regions which indeed grow with time as illustrated in Fig.~\ref{fig3}.
See Section \ref{resu} for more details.
}
\label{fig5}
\end{center}
\end{figure}

Finally, in Fig.~\ref{fig4} we show the peculiar velocities of the two components.
The left panel shows the peculiar velocities of the $M$ component with respect to the background, which are given by $v_{M} \equiv Y (H_{A}-H_{\rm out})$. This definition again relies on the fact that $r_{p} \simeq Y$.
As expected these peculiar velocities are growing with time as the underdensity becomes emptier and emptier (see also the left panel of Fig.~\ref{fig1}).
Opposite is, however, the case of the peculiar velocities of the $N$ component with respect to the $M$ component (right panel of Fig.~\ref{fig4}), which decrease with time: the gravitational attraction between the two components dumps their relative velocities.
The plots of Fig.~\ref{fig4} are with respect to the ``reference frame'' of the background and of the center of the underdensity, as the velocities go to zero at $r=0$ and $r= r_{b}$.
In Fig.~\ref{fig5} we re-plot the same curves, but with the velocity of the peak in density subtracted. Fig.~\ref{fig5} therefore better shows the velocity field close to the overdensity, which appears to be  natural, as matter is falling towards the peak in the density from both directions.

Finally, this model maintains the scale invariance valid for the simpler LTB model~\cite{Marra:2007pm} if the initial peculiar velocities are properly scaled: $ \tilde v_{c}(r,\bar t) / \tilde r_{b}= v_{c}(r,\bar t) /r_{b}$ where tilde marks the quantities relative to an inhomogeneous patch of different radius.

\section{Collapse in the presence of dark energy with negligible speed of sound} \label{cde}

As a second application of our general model we will consider an inhomogeneous sphere embedded into a flat $w$CDM universe.
The inhomogeneities will be given by a pressureless matter component and by a dark energy source with constant equation of state $w_{\rm out}<-1/3$.
This solution becomes the usual $\Lambda$LTB model if $w=w_{\rm out}=-1$ and the Lema\^{i}tre model (possibly in non comoving coordinates) if the density of the dust component is set to zero.

We will use the reference frame comoving with the matter component, which we label with $M$, so that $\lambda=0$ and Eqs.~(\ref{consam}-\ref{dustevo}) hold.
We will label the dark energy component with ``$w$'' and use the simpler notation $w_{w}=w$, $p_{w}=p$, $\gamma_{w}=\gamma$, $v_{w,\, p}=v_{p}$, $v_{w,\, c}=v_{c}$ as the corresponding $M$ quantities are trivial.
As we are discussing a fluid with negative pressure we cannot consider adiabatic perturbations for which $w(r,t) = w_{\rm out}$: this would indeed give a negative speed of sound ($c_{s}^{2}=w_{\rm out}$) which would cause an unstable growth of perturbations~\cite{Hu:2004kh}.
We will consider instead the following $r$-dependent non-adiabatic equation of state:
\begin{equation}
w(r,t) = w_{\rm out} \left [{\rho_{w}(r,t) \over  \rho_{w}^{\rm out} (t)} \right ]^{\alpha -1} \,,
\end{equation}
which implies the following speed of sound:
\begin{equation}
c_{s}^{2}= {\partial p_{w} \over \partial \rho_{w}} =  \alpha \, w(r,t)  \,, \label{cs}
\end{equation}
so that for $\alpha \le 0$ the speed of sound is positive (the adiabatic speed of sound is recovered with $\alpha =1$, see e.g. Ref.~\cite{Kolb:1990vq}).

Particularly interesting is the case $\alpha=0$, for which the pressure gradients are absent and the speed of sound vanishes, with the consequence that matter and dark energy collapse together following the geodesic flow.
The possibility of a dark energy fluid with negligible speed of sound has been investigated in the literature under various assumptions, and generally requires a non canonical scalar field like $k$-essence, as  the standard quintessence models with canonical scalar fields always have $c_{s}=1$ (see, for example, \cite{Creminelli:2008wc, Bertacca:2008uf, Lim:2010yk,Bertacca:2010ct,Li:2011sd} and references therein).
Apart from theoretical considerations, the fact that the speed of sound of dark energy may vanish opens up new observational consequences. Indeed, the absence of dark energy pressure gradients allows instabilities to develop on all scales, also on the small scales where dark matter perturbations become non-linear.  We expect, therefore, dark energy to modify in a detectable manner the growth history of dark matter halos not only through its different background evolution but also by actively participating to the structure formation process, in the linear and non-linear regime~\cite{Creminelli:2009mu, Sanchez:2010ng, Sefusatti:2011cm,Anselmi:2011ef}.
The observational consequences of a clustering dark energy have been indeed extensively studied. See, for example, \cite{DeDeo:2003te,Amendola:2003bz,Weller:2003hw,Bean:2003fb,Hu:2004yd,Hannestad:2005ak,Corasaniti:2005pq, Bertacca:2011in} for the impact  on large-scale structures and cosmic microwave background.

Finally, we would like to point out that, as we will use the matter reference frame, the speed of sound given in Eq.~(\ref{cs}) will not be in general in the dark energy rest frame, which is usually used in order to have a gauge independent expression for the speed of sound.
However, as we will not introduce initial peculiar velocities, if $c_s=0$ then the matter and dark energy rest frames will always coincide because of the absence of pressure gradients. The latter will also be approximately true for the case of nonzero but negligible speed of sound as the peculiar velocities developed by the dark energy fluid will be very small (see the right panel of Fig.~\ref{fig9}).

\subsection{Initial and boundary conditions}

We fix the background model by setting $h=0.7$ and $\Omega_{M0}=0.3$ as in Section \ref{colla}.
About the equation of state we will study the following three cases: $w_{\rm out}=-0.8$ and $\alpha =0$ in Section \ref{evo0}, $w_{\rm out}=-0.8$ and $\alpha =-10^{-10}$ in Section \ref{evonz} and $w_{\rm out}=-1.2$ and $\alpha =-10^{-10}$ in Section \ref{evoph}.
The inhomogeneous sphere will have a comoving radius of $r_{b}=10 h^{-1}$~Mpc and we will give initial conditions at the time $\bar t$ corresponding to the background redshift $\bar z=1000$.
We set again the scale function at initial time to $Y(r, \bar t)= a(\bar t) r$.

We have to give initial conditions for $F(r, \bar t)=\bar F_{M}(r)+\bar F_{w}(r)$.
We model the initial matter inhomogeneity as:
\begin{equation}
{\bar F_{M}(r) \over \bar F_{M}^{\rm out}(r)} = 1+ \delta_{M}(\bar t) \, \frac{1-\tanh ( (r-r_0 - r_{\rm offset}) / 2\Delta r )}{1+\tanh (r_0/ 2\Delta r )} \,,
\end{equation}
where $\bar F_{M}^{\rm out}= {4 \pi \over 3}  a^{3}(\bar t) r^{3} \rho_{M}^{\rm out} (\bar t)$ is the background gravitating mass, $\delta_{M}$ is the density contrast, and the parameters $r_{0}$ and $\Delta r$ characterize respectively size and steepness of the density profile.
The parameter $r_{\rm offset}$ is used to translate the profile in order to have a smooth origin. We will adopt the following values: $r_{0}=r_{\rm offset}=0.25 \, r_{b}$, $\Delta r = 0.3 \, r_{0}$ and $\delta_{M}(\bar t) = 1.5 \cdot 10^{-3}$, that is, we have a central overdensity.
The initial matter density is then:
\begin{equation}
\rho_{M}(r,\bar t)= {\bar F_{M}'(r)  \over 4 \pi a^3(\bar t) r^2 } \,.
\end{equation}

Next, we can find the initial curvature function $E$ by demanding that the universe has the same age $\bar t$ for any $r$, i.e., from Eq.~(\ref{hbb}):
$t_{A}(r,\bar t) = \bar t $.
As $\bar F_{w}(r) \ll \bar F_{M}(r)$ at $\bar t$, this condition will be independent of the dark energy initial conditions. In particular at the initial time the model is very close to a (dust) LTB model, so that it is possible to find $E$ analytically by means of the following expression valid for $\delta_{M} \ll 1$~\cite{VanAcoleyen:2008cy}:
\begin{equation}
E(r, \bar t) \simeq  - {5\over 6}  \Big (a(\bar t)  H_{\rm out}(\bar t) \, r \Big)^2  \left({\bar F_{M}(r) \over \bar F_{M}^{\rm out}(r)} -1 \right) \,.
\end{equation}
These initial conditions are free from decaying modes~\cite{Zibin:2008vj}.

Finally, we take the dark energy component to have no initial peculiar velocities, which means that it will have, with respect to the background, the peculiar velocities $Y \Delta H$ of the matter component.
Therefore, in order to  have initial conditions without decaying modes in the dark energy component, we have to set the initial profile as:
\begin{equation}
{\bar F_{w}(r) \over \bar F_{w}^{\rm out}(r)} = 1+\delta_{w}(\bar t) \, \frac{1-\tanh ( (r-r_0 - r_{\rm offset}) / 2\Delta r )}{1+\tanh (r_0/ 2\Delta r )} \,,
\end{equation}
where $\bar F_{w}^{\rm out}= {4 \pi \over 3}  a^{3}(\bar t) r^{3} \rho_{w}^{\rm out} (\bar t)$ and initial contrast is:
\begin{equation}
\delta_{w}(\bar t) ={1+w_{\rm out} \over 1-3 w_{\rm out}} \; \delta_{M}(\bar t) \,,
\end{equation}
which is valid during matter domination for sub-Hubble perturbations and $c_{s}^{2}\ll1$~\cite{Ballesteros:2010ks}.
The initial dark energy density is then:
\begin{equation}
\rho_{w}(r,\bar t)= {\bar F_{w}'(r)  \over 4 \pi a^3(\bar t) r^2 } \,.
\end{equation}
With this we have all the necessary initial conditions.

\subsection{Evolution for zero speed of sound} \label{evo0}

\begin{figure}
\begin{center}
\includegraphics[width= .49\textwidth,height=6.15 cm]{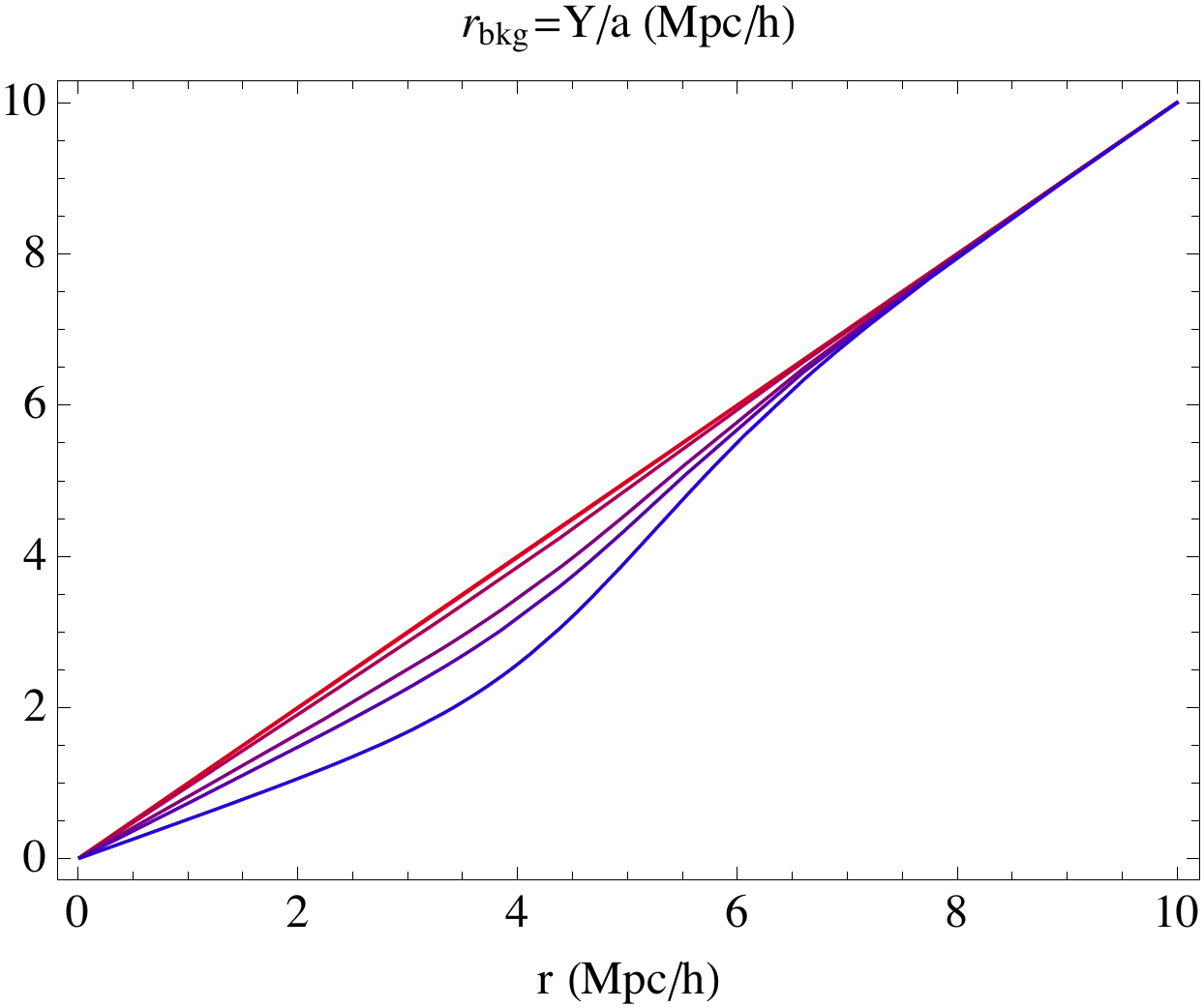}
\includegraphics[width= .50\textwidth]{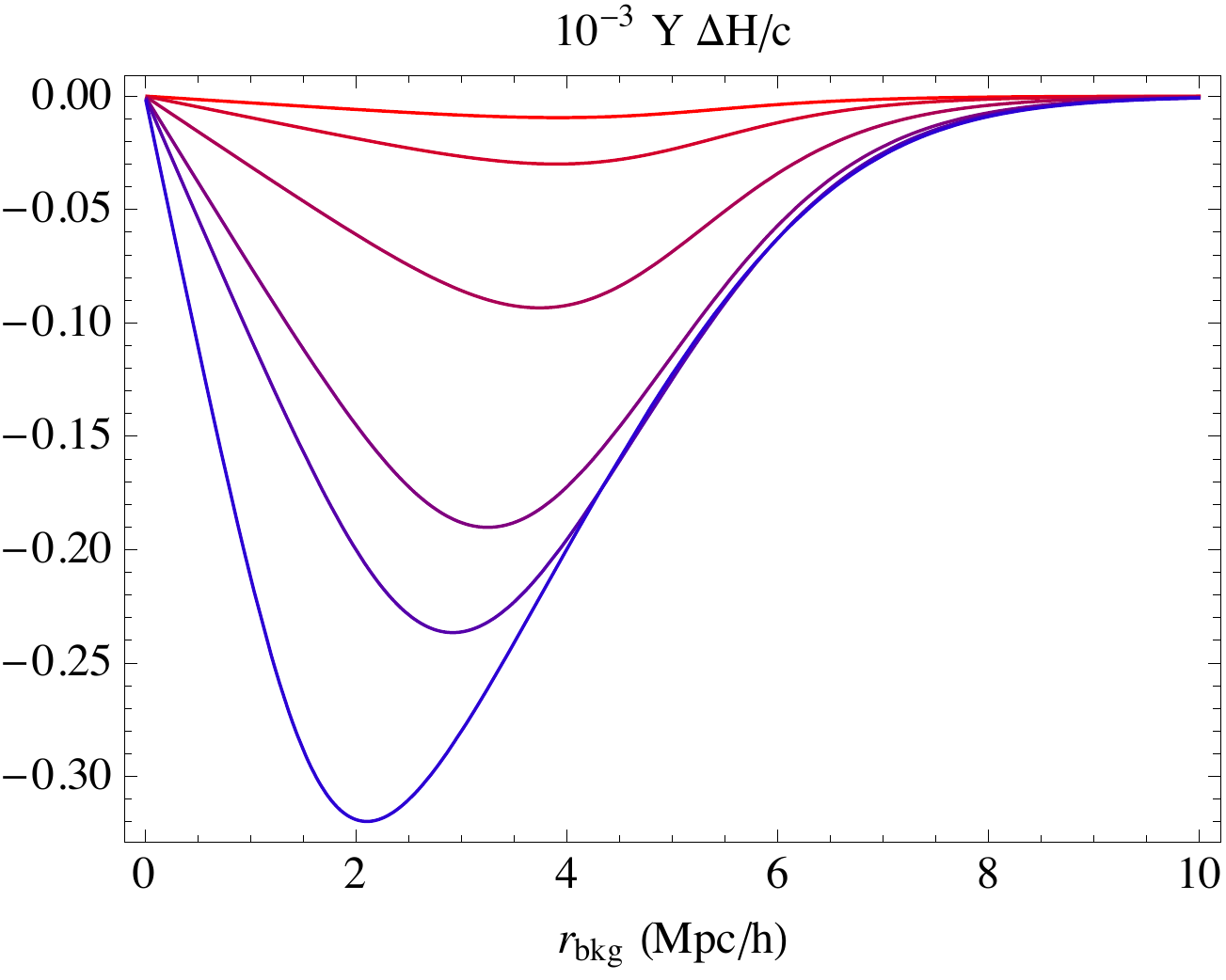}
\caption{
Left panel: evolution of the scale function $Y$ for the times corresponding to the background redshifts of $z=1000$ (red), 100, 10, 2, 1, and 0 (blue). In particular, this plot shows the relation between the coordinate radius $r$ and the background comoving radius $r_{\rm bkg}=Y/a$ which will be used in the other plots.
Right panel: evolution of the peculiar velocities $v_{M}\equiv Y (H_{A}-H_{\rm out})$ of the $M$ component with respect to the background.
Velocities are directed inwards and increase in magnitude with time as the central overdensity becomes denser.
See Section \ref{evo0} for more details.
}
\label{fig6}
\end{center}
\end{figure}

The left panel of Fig.~\ref{fig6} shows the evolution of the scale function $Y$  for $w_{\rm out}=-0.8$ and $\alpha =0$: in the interior region the scale function is expanding slower because of the more matter and consequently positive curvature present.
All the other figures will be plotted with respect to the background comoving radius defined as $r_{\rm bkg}=Y(r,t)/a(t)$, which as explained in Section~\ref{resu} is a good definition of distance.
The right panel shows the peculiar velocities of the $M$ component with respect to the background.
As there are no initial decaying modes, the peculiar velocities are natural, with the matter falling towards the peak in the density, as shown by the left panel of Fig.~\ref{fig6}. Moreover, the velocities grow with time as the central overdensity becomes denser and denser.

\begin{figure}
\begin{center}
\includegraphics[width= .49\textwidth, height=6.35 cm]{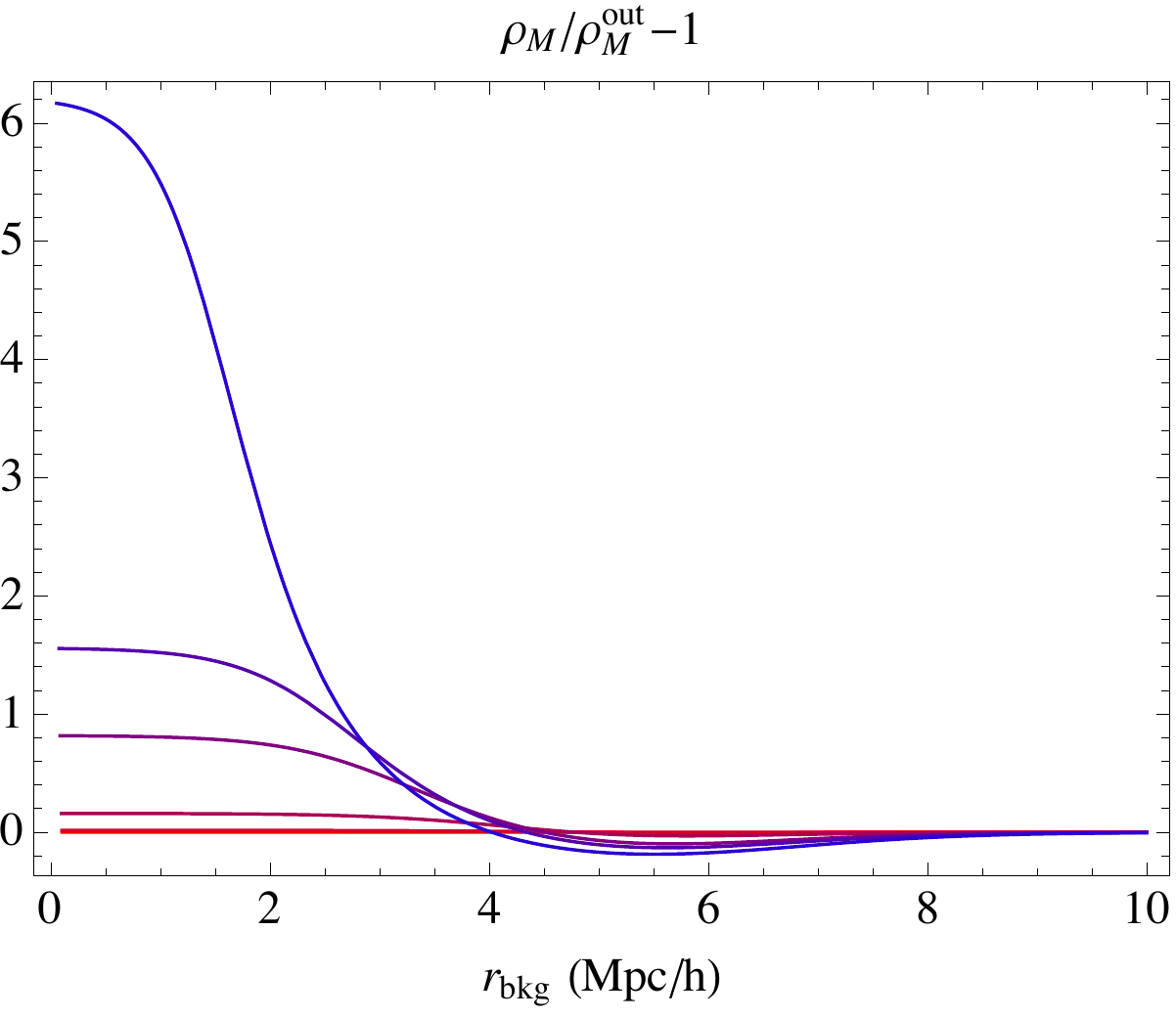}
\includegraphics[width= .50\textwidth]{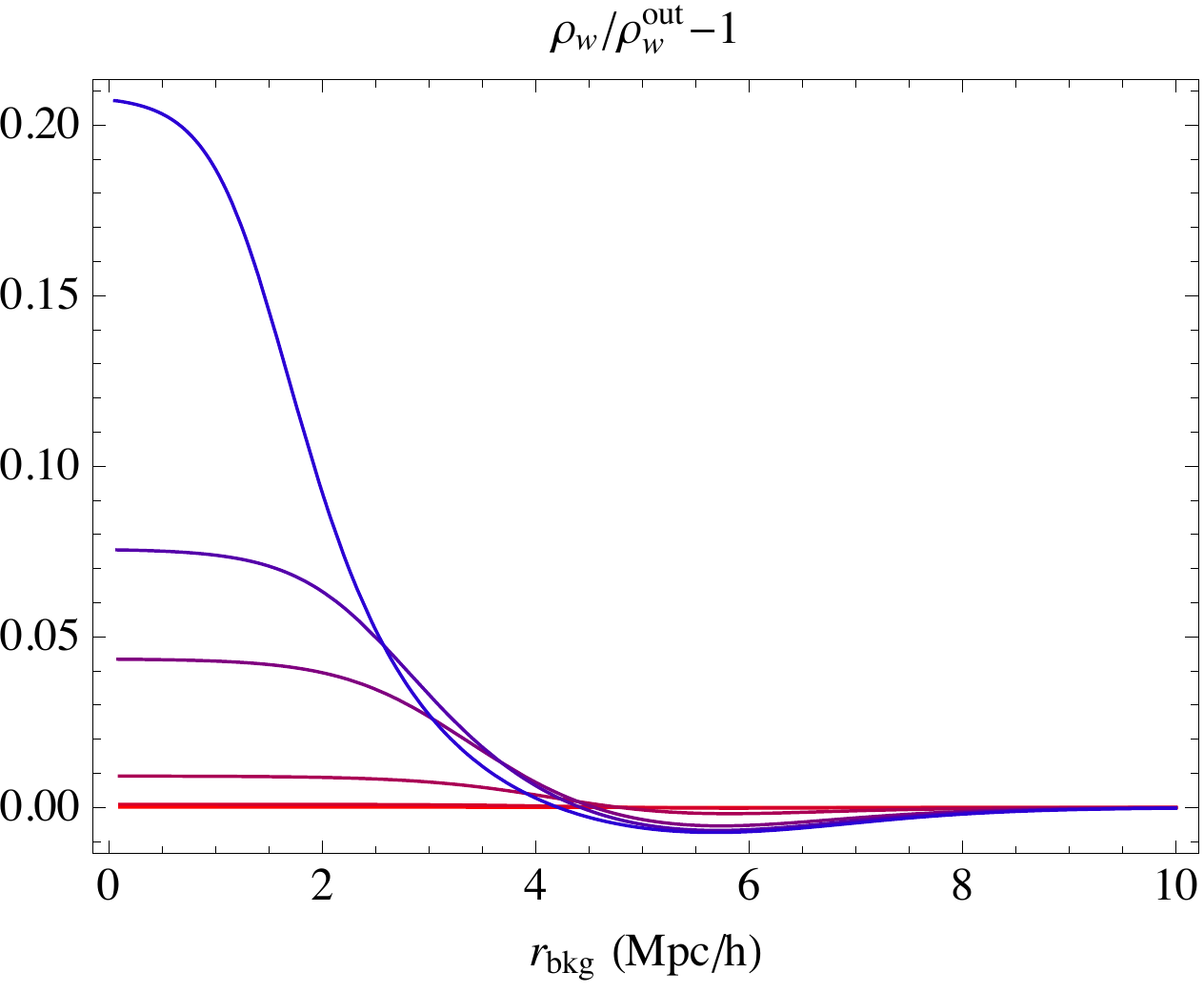}
\caption{
Evolution of the matter contrast (left panel) and dark-energy contrast (right panel) for $z=1000$ (red), 100, 10, 2, 1, and 0 (blue).
Both the density contrasts show a realistic evolution, with the overdense region becoming denser and denser.
Note, in particular, how the dark energy component follows the evolution of the matter component, with the difference of a slower growth rate due to the nonzero pressure.
See Section \ref{evo0} for more details.
}
\label{fig7}
\end{center}
\end{figure}

\begin{figure}
\begin{center}
\includegraphics[width= .49\textwidth]{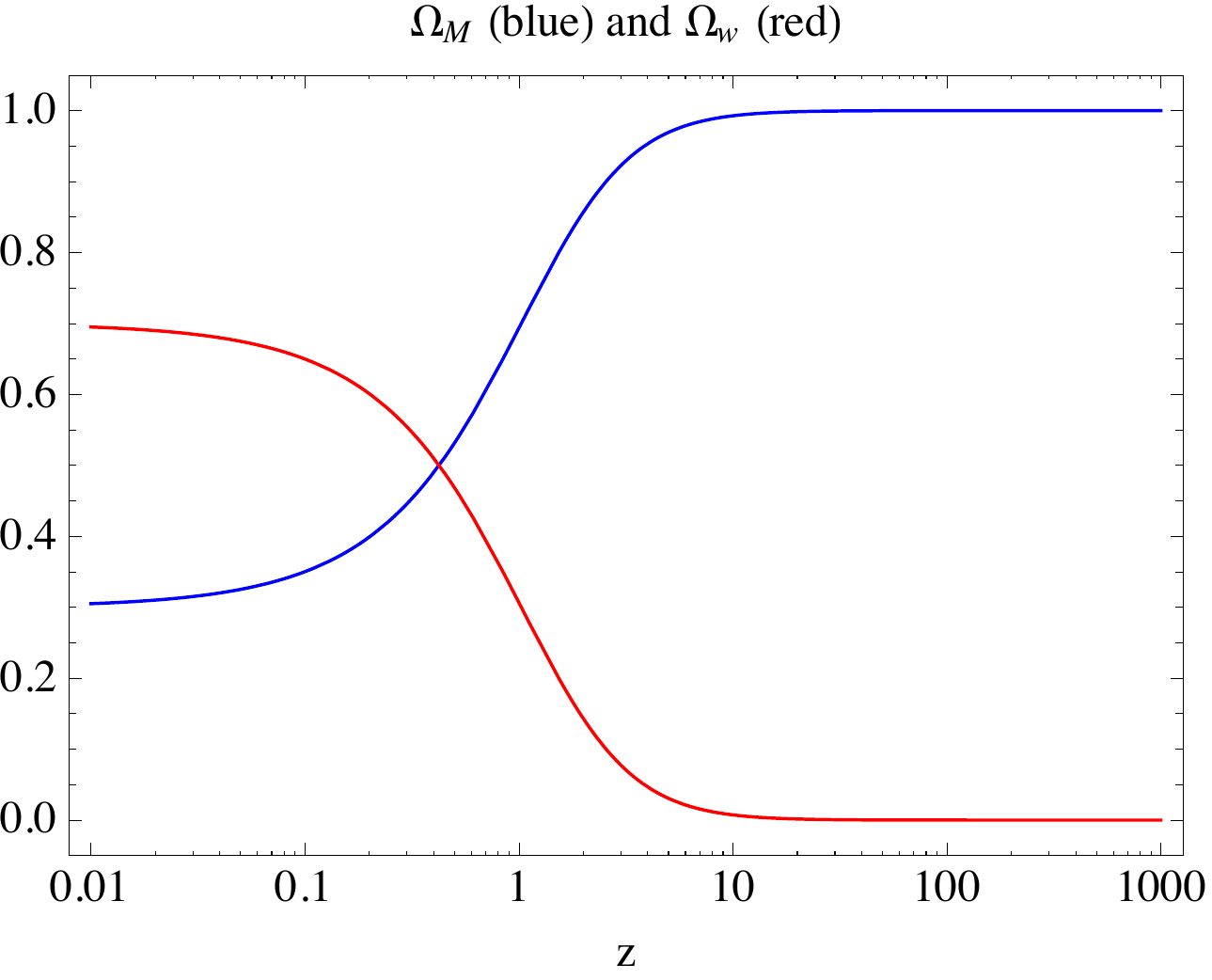}
\includegraphics[width= .49\textwidth]{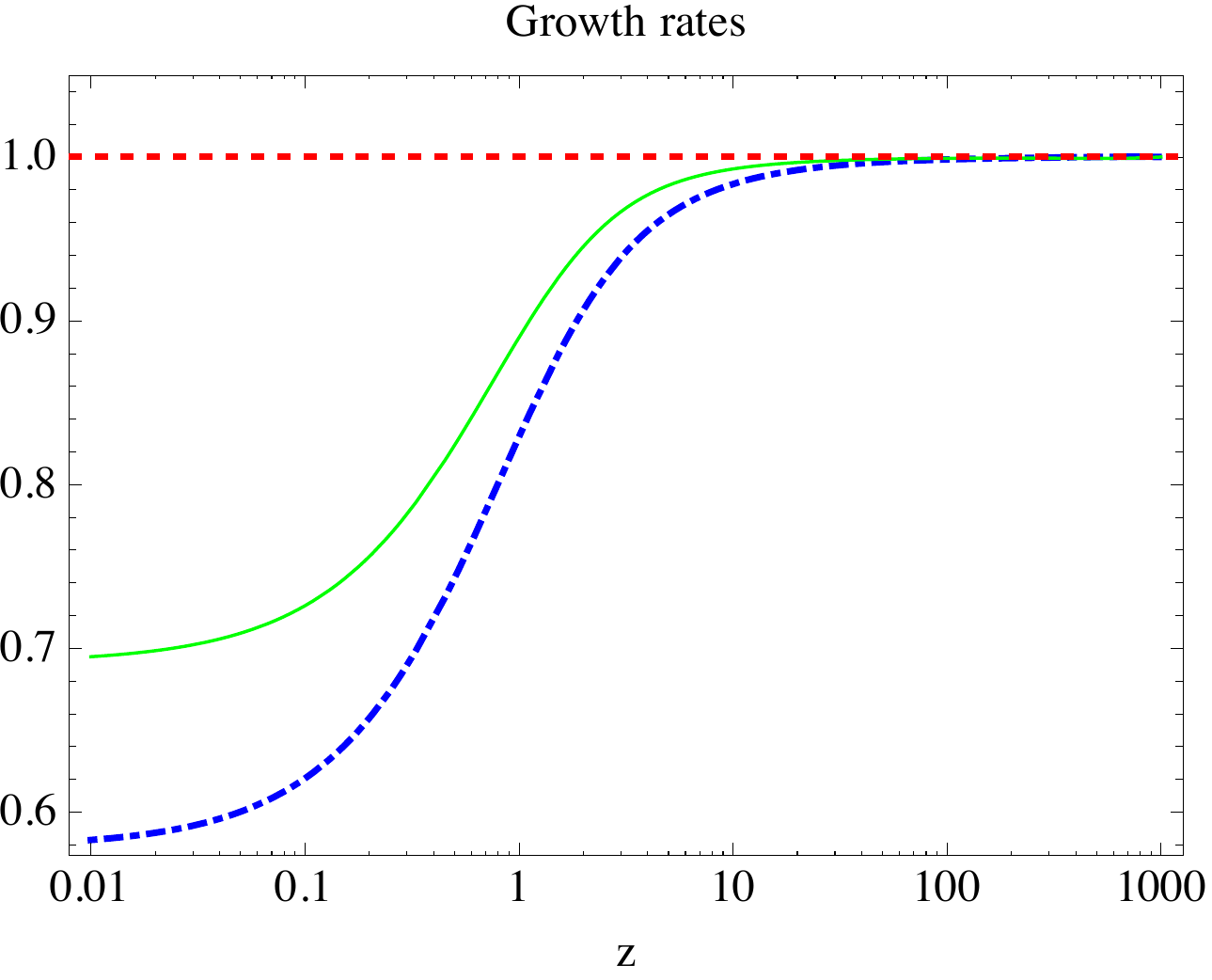}
\caption{
Left panel: redshift evolution of the matter and dark energy density parameters.
Right panel: evolution of the ratio of the contrasts, $\delta_{w}/\delta_{M}$, normalized to the value valid during matter domination, ${1+w_{\rm out} \over 1-3 w_{\rm out}}$.
The blue (dot-dashed) line is relative to $r_{\rm bkg}=1 h^{-1}$Mpc and the green (solid) line to $r_{\rm bkg}=6 h^{-1}$Mpc, see Fig.~\ref{fig7}.
See Section \ref{evo0} for more details.
}
\label{fig8}
\end{center}
\end{figure}

In Fig.~\ref{fig7} the evolution of the density contrasts of the matter and dark energy components is shown.
The evolution, as remarked above, is realistic. In particular the dark energy component closely follows the evolution of the matter component, as one should expect from a fluid with vanishing speed of sound.
Because of the absence of pressure gradients peculiar velocities between the two components never develop (if initially zero) and the two fluids both follow the same geodesic flow.\footnote{For the same reason the curvature is time independent, as it is easy to see from Eq.~(\ref{fset2}).}
As it is clear from the plot, however, the growth rates are different. This has to be expected because the negative pressure generally slows down the evolution of perturbations.
To be quantitative, we have plotted in right panel of Fig.~\ref{fig8} the evolution of the ratio of the contrasts, $\delta_{w}/\delta_{M}$, which we normalized to the value valid during matter domination, ${1+w_{\rm out} \over 1-3 w_{\rm out}}$.
As one can see, the agreement is perfect when $\Omega_{M}\sim 1$ (see left panel of Fig.~\ref{fig8}), showing that the chosen initial conditions are indeed free from decaying modes.
During the nonlinear dark-energy--dominated phase the ratio departs as expected from the matter-domination prediction, and in particular becomes different in overdense and underdense regions, as illustrated in the right panel of Fig.~\ref{fig8} where the blue line is relative to $r_{\rm bkg}=1 h^{-1}$Mpc (overdensity) and the green line to $r_{\rm bkg}=6 h^{-1}$Mpc (underdensity). This could be due to the fact that even though the two components have similar profiles, in the overdense region the matter component reaches the nonlinear evolution earlier (the dark energy is actually  almost linear at the present time).

It is also interesting to compare the evolution to the case of no dark energy. We find that the matter perturbation grows to a maximum contrast of about $\sim90$.
Alternatively, we find that an initial perturbation of $\delta_{M}(\bar t) = 1.1 \cdot 10^{-3}$ grows to the same present-day contrast as the initial perturbation of $\delta_{M}(\bar t) = 1.5 \cdot 10^{-3}$ studied in this Section.

Finally, let us comment on the fact that because the speed of sound is zero, there are no characteristic length scales associated to the dark energy clustering and the spherical collapse remains independent of the size of the object~\cite{Creminelli:2009mu}.
The characteristic length scale associated to the dark energy clustering is indeed the sound horizon scale, $L_{s}= a \int c_s \, dt/ a$, which vanishes for $c_s=0$ so that clustering takes place on all scales.
In other words, this model maintains the scale invariance valid for the simpler LTB model~\cite{Marra:2007pm}.

\subsection{Evolution for \boldmath $c_{s} \sim 10^{-5}$} \label{evonz}

\begin{figure}
\begin{center}
\includegraphics[width= .49\textwidth]{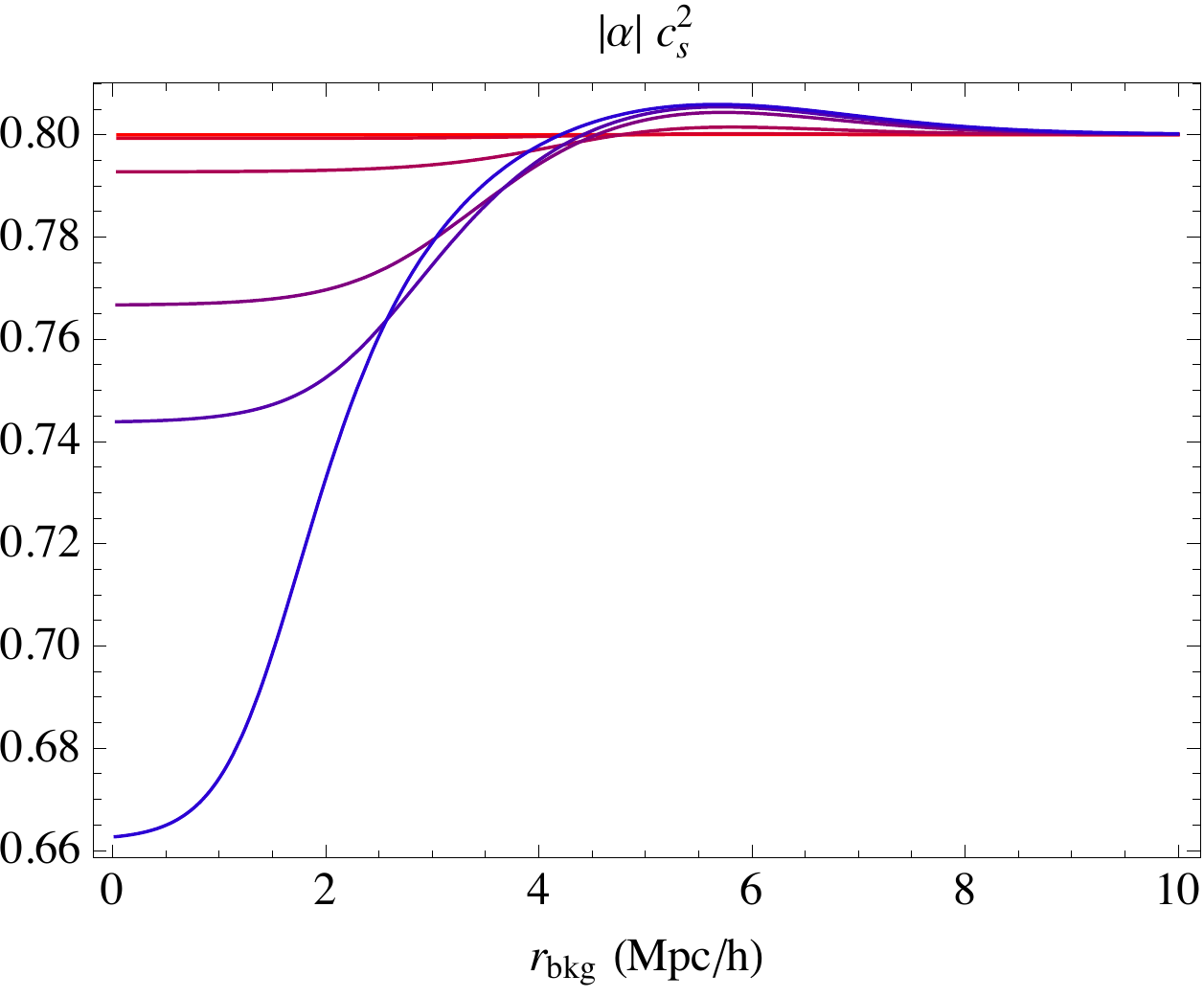}
\includegraphics[width= .50\textwidth]{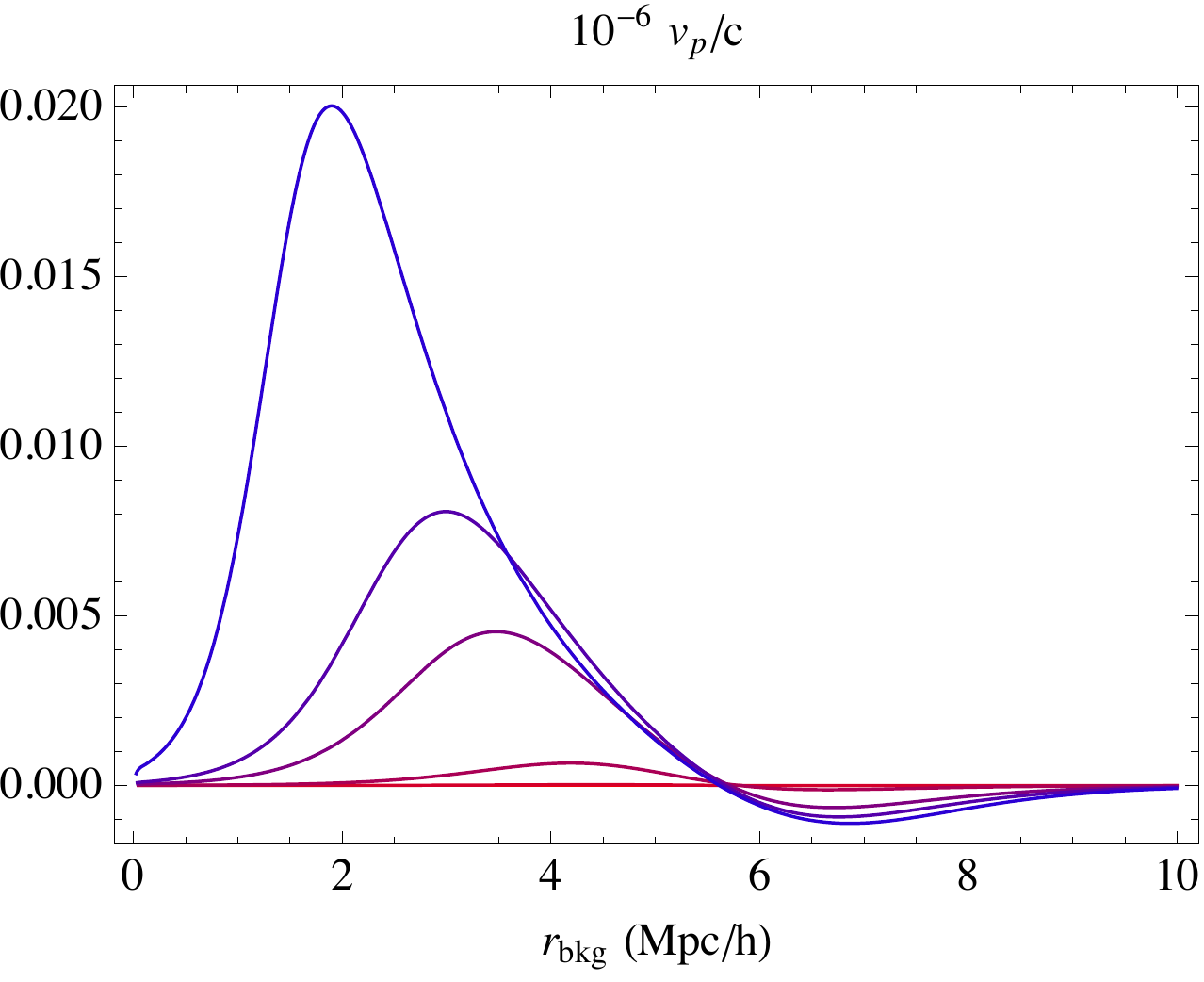}
\caption{
Left panel: evolution of the speed of sound $c_{s}^{2}$ for the times corresponding to the background redshifts of $z=1000$ (red), 100, 10, 2, 1, and 0 (blue).
Right panel: evolution of the peculiar velocities of the dark-energy component with respect to the matter component.
See Section \ref{evonz} for more details.
}
\label{fig9}
\end{center}
\end{figure}

We will now consider the model of the previous Section ($w_{\rm out}=-0.8$) with $\alpha=-10^{-10}$, which corresponds to a speed of sound of order $c_{s} \sim 10^{-5}$, as one can see from the left panel of Fig.~\ref{fig9}.
This latter value implies a sound horizon of order $L_{s} \sim 80 h^{-1}$ kpc, much less than the inhomogeneity scale considered in this example, so that the overall dynamical evolution shown in Figs.~\ref{fig6}-\ref{fig8} is basically unchanged.
The new features of this case are pressure gradients and peculiar velocities. In order to understand their behavior it is useful to work out the following approximations valid for $|\delta_{w}|, |\alpha| \ll 1$:
\begin{equation}
c_{s}^{2} \simeq \alpha \, w_{\rm out} (1- \delta_{w}) \,, \qquad \quad
\delta_{p} \equiv {p - p^{\rm out} \over |p^{\rm out}|} \simeq -\alpha \, \delta_{w} \,.
\end{equation}
From the previous relations it is easy to understand how speed of sound and pressure gradients are related to the density contrast. In particular it is $\delta_{p}' \sim c_{s}' \sim - \delta_{w}'$, as one can also see by comparing the left panel of Fig.~\ref{fig9} with the right panel of Fig.~\ref{fig7}.
As a consequence, the pressure gradients will push the dark energy component out of the free-falling geodesic away from the overdense regions, as shown in the right panel of Fig.~\ref{fig9}.
One can indeed obtain at linear order (see, e.g., \cite{Bjaelde:2010qp}) the approximate relation $\dot v_{p} \sim - {c_{s}^{2} \over 1+w_{\rm out}} \, \delta_{w}' $.

\subsection{Evolution for \boldmath $w< -1$} \label{evoph}

\begin{figure}
\begin{center}
\includegraphics[width= .49\textwidth,, height=6.15 cm]{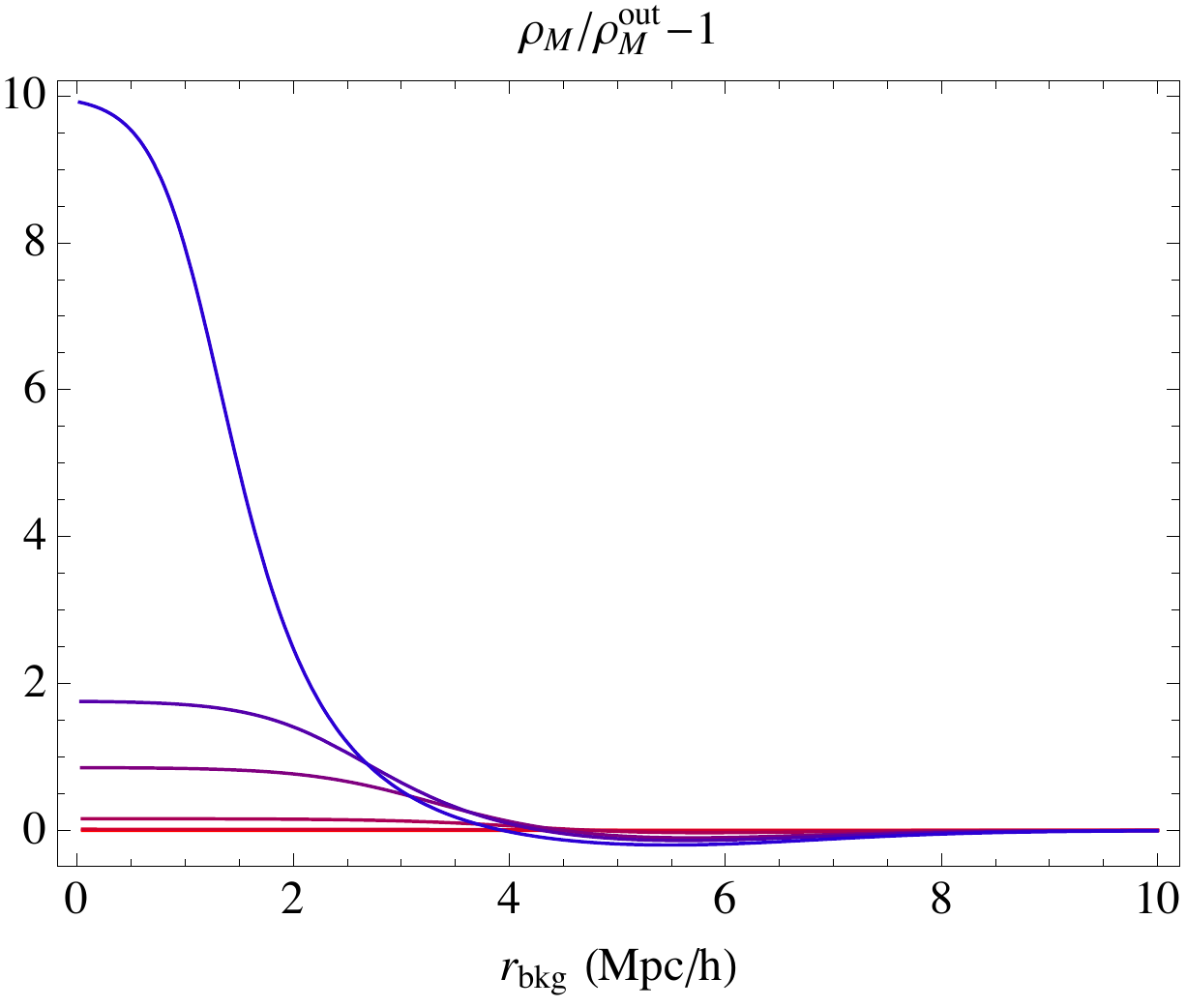}
\includegraphics[width= .49\textwidth]{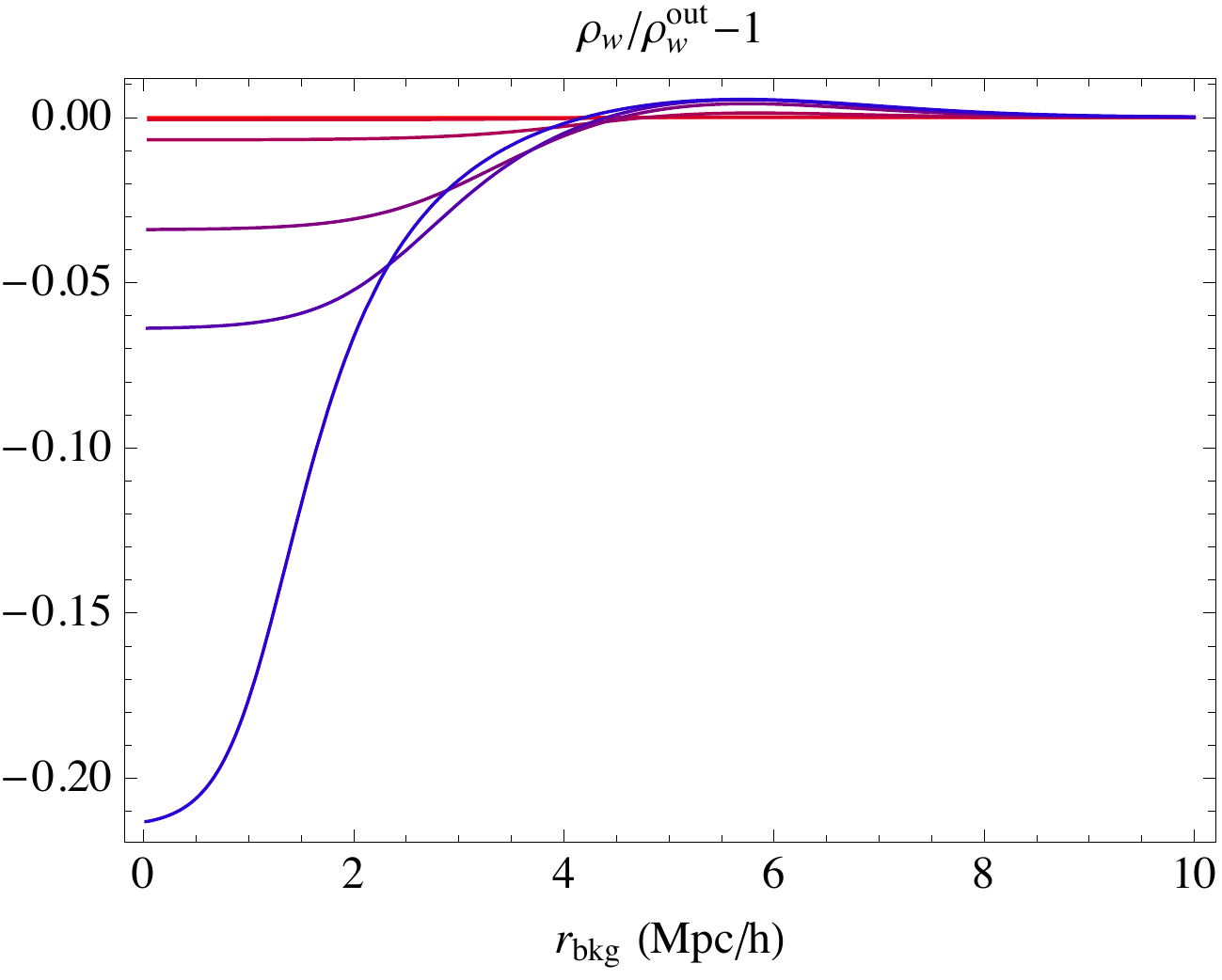}
\caption{
Evolution of the matter contrast (left panel) and dark-energy contrast (right panel) for $z=1000$ (red), 100, 10, 2, 1, and 0 (blue).
Note that the evolution of the dark energy component still follows the matter flow but develops an underdensity instead of an overdensity. 
See Section \ref{evoph} for more details.
}
\label{fig10}
\end{center}
\end{figure}

\begin{figure}
\begin{center}
\includegraphics[width= .49\textwidth,, height=6.15 cm]{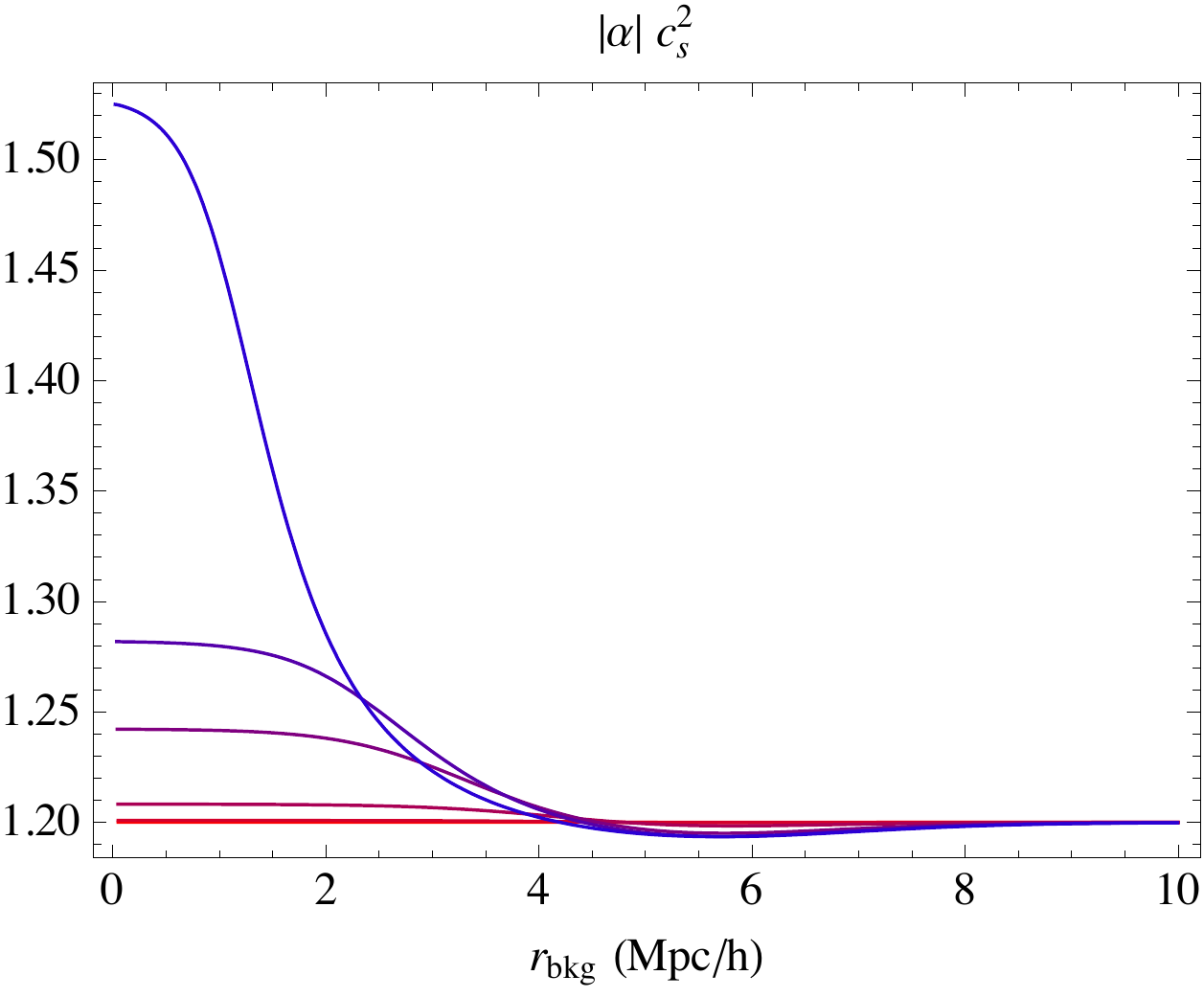}
\includegraphics[width= .49\textwidth]{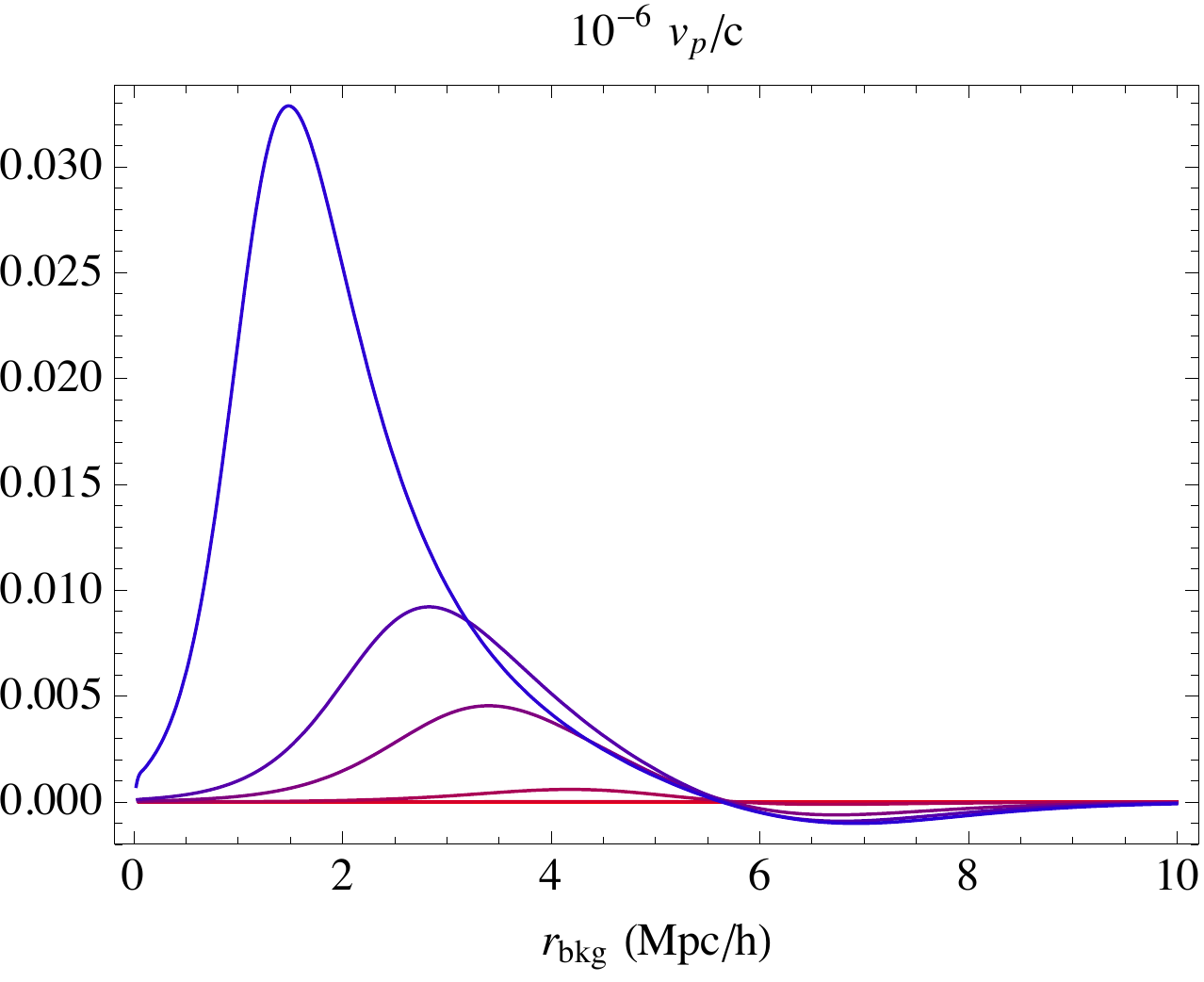}
\caption{
Left panel: evolution of the speed of sound $c_{s}^{2}$ for the times corresponding to the background redshifts of $z=1000$ (red), 100, 10, 2, 1, and 0 (blue).
Right panel: evolution of the peculiar velocities of the dark-energy component with respect to the matter component.
See Section \ref{evoph} for more details.
}
\label{fig11}
\end{center}
\end{figure}

Finally, we will now consider the case of $w_{\rm out}=-1.2$ and $\alpha =-10^{-10}$.
We chose the same order of magnitude for the (positive) speed of sound of the previous Section for comparison sake, see the left panel of Fig.~\ref{fig11}. Clearly, we can decrease the magnitude of $\alpha$ in order to prevent possible pathologies~\cite{Creminelli:2008wc}.

As one can see from the left panel of Fig.~\ref{fig10}, the evolution of the matter component is similar to the previous case, with a moderately higher contrast at late times probably due to the different expansion history: for $w_{\rm out}=-1.2$ the dark-energy dominance happens indeed later with respect to the case for $w_{\rm out}=-0.8$.\footnote{Similar considerations apply to the case of a cosmological constant which gives a present-day contrast of about $\sim8$, roughly half way between the cases studied in Sections \ref{evo0} and \ref{evoph}.}
Quite different is instead the evolution of the dark energy component, which still follows the matter flow but develops an underdensity rather than an overdensity (see the right panel of Fig.~\ref{fig10}). This can be understood from the fact that the relation $\delta_{w}/\delta_{M} \sim {1+w_{\rm out} \over 1-3 w_{\rm out}}$ changes sign for $w_{\rm out} < -1$.
It is interesting to note that a similar behavior was found in the analysis of Ref.~\cite{Sanchez:2010ng} of perturbations in scalar-tensor cosmologies, where the non-minimal coupling of the field is indeed responsible for the crossing of the so-called phantom divide ($w_{\rm out}=-1$).
As one can see from the right panel of Fig.~\ref{fig11}, the evolution of the peculiar velocities is instead approximately the same one of the previous Section, as in the relation $\dot v_{p} \sim - {c_{s}^{2} \over 1+w_{\rm out}} \, \delta_{w}'$ both $\delta_{w}'$ and $1+w_{\rm out}$ change sign.

\section{Comparison with previous work on exact solutions} \label{litera}

In this Section we will compare our findings to previous work dealing with exact solutions.
As explained in Section \ref{intro} the model here presented is a generalization to the case of $n$ decoupled and non-comoving perfect fluids of the Lema\^{i}tre model \cite{Lemaitre:1933gd} (see \cite{Bolejko:2005tk, Lasky:2006mg, Alfedeel:2009ef, Lasky:2010vn} for recent contributions) which describes the dynamics of a spherically-symmetric perfect fluid.
Therefore, the model of Eqs.~(\ref{fset1}-\ref{fset5}) straightforwardly reduces to the Lema\^{i}tre model, possibly with the addition of a cosmological constant.

In the case of one source only, our equations can be used to describe the fluid in non-comoving possibly-geodesic coordinates, which may be desirable if the fluid has pressure gradients and its four-velocity undergoes acceleration.
Within our formalism this is accomplished by simply taking a geodesic reference-frame four-velocity $u_{\rm rf}^{\alpha}$, i.e., by  setting $\lambda'=0$ (see Eq.~(\ref{exparf1})).
This can be useful to disentangle the effect of pressure gradients on the evolution of the model as shown, for instance, by Eq.~(\ref{kdot1}) for the time evolution of the curvature function. In the latter equation, indeed, only the second term on the right-hand side contributes if comoving coordinates are used and only the first term on the right-hand side contributes if geodesic coordinates are used.
Spherical models have been discussed previously in the literature in non-comoving coordinates, albeit with a different purpose.
The main idea was indeed that an exact solution with a simple appearance in non-comoving coordinates may become extremely complex when transformed to the appropriate comoving system, or that the integration of the partial differential equations necessary to obtain the comoving coordinates may even represent an intractable mathematical problem.
We refer the interested reader to \cite{McVittie1977, Bonnor1993, Knutsen95, Herrera:2002im, Davidson2003, Ishak:2004dd, Herrera:2011kd} and to \cite{Stephani:2003tm} and references therein.

Another line of research focused on interpreting two or more fluid components as a single effective fluid, see for example \cite{AndrzejKrasinski:1997zz} and references therein.
The work of \cite{letelier80, letelier86} showed indeed that a mixture of fluids whose four-velocities lay on a two-plane can always be reinterpreted as a single anisotropic fluid. 
The model of this paper satisfies this last requirement and it would be interesting to investigate in future work the amount of anisotropic pressure generated by the evolution of a cosmological multi-component fluid modeling, e.g., baryons, dark matter and dark energy. As a simple example we can apply here the work of \cite{letelier80} to the model of Section \ref{colla} featuring two non-comoving dust components in a flat $\Lambda$CDM universe.
Following \cite{letelier80}, the energy-momentum tensor of the effective fluid is:
\begin{equation}
T^{\alpha\beta}_{\rm eff}  =  \rho_{\rm eff} \, u_{\rm eff} ^{\alpha}u_{\rm eff} ^{\beta} +  S^{\alpha\beta}_{\rm eff}   \,,
\end{equation}
and a short calculation shows that the rest energy density is:
\begin{equation}
\rho_{\rm eff} = {1 \over 2}(\rho_M + \rho_N) + {1 \over 2} \sqrt{(\rho_M + \rho_N)^2 + 4 \rho_M  \rho_N \, ( \gamma^2-1)} \,,
\end{equation}
and that the components of the diagonalized stress tensor are:
\begin{eqnarray}
p_A&=& 0 \,,\\
p_R &=& -{1 \over 2}(\rho_M + \rho_N) + {1 \over 2} \sqrt{(\rho_M + \rho_N)^2 + 4 \rho_M  \rho_N \, (\gamma^2-1)} \,,
\end{eqnarray}
where $p_A$ and $p_R$ are the angular and radial pressures, respectively.
Clearly,  for $\gamma \rightarrow 1$ ($v_c \rightarrow 0$) the effective fluid becomes simply the sum of the two fluids and $p_A=p_R=0$. For a nonzero radial velocity $v_c \neq 0$, however, the radial pressure is positive $p_R>0$ and the effective fluid is anisotropic. See \cite{letelier80} for more details.
Finally, we would like to mention that the authors of \cite{letelier86} claim that the opposite process is also feasible, i.e., that given a particular anisotropic fluid, it is always possible to find a multi-fluid model with an equivalent energy-momentum tensor.
Note, however, that this last claim was argued not to be valid for a general anisotropic fluid~\cite{leon87} such as the anisotropic fluid models of compact objects like neutron stars~\cite{bowers74, cosenza82}.

Finally, \cite{gr-qc/9907010} studied anisotropic pressure in the context of the LTB metric in order to investigate the evolution of inhomogeneities in an interacting photon-baryon mixture.
A physically-plausible hydrodynamical description of cosmological matter in the radiative era between nucleosynthesis and decoupling was given, emphasizing its thermodynamical consistency. It would be interesting to develop such considerations also in the presence of a dark matter component using the multi-fluid model presented in this paper.

\section{Conclusions} \label{conclusions}

We have extended the Lema\^{i}tre model to the case of $n$ decoupled and non-comoving perfect fluids with general equations of state.
We have expressed the full set of $3+2n$ exact equations governing the dynamics of the model in a concise and transparent way thanks to the use of physically meaningful quantities like, for example, expansion and acceleration rates, expansion and acceleration scalars, total derivatives along four-velocity and acceleration.
This general solution can have many possible applications, both at early and late times.

At late time, one can study the evolution of overdensities and underdensities in a universe where dark energy is not the cosmological constant and, in particular, can be inhomogeneous.
This possibility may also induce an inhomogeneous variation of fundamental constants if a coupling between the dark energy and the matter-radiation Lagrangian is allowed.
Moreover, dark matter and baryons can be described as two separate fluids, the latter possibly featuring pressure.
Pressure in general can have a non-negligible effect on the cosmological models and this should be taken into consideration while interpreting cosmological datasets.
At early times the contribution of radiation can be included, which may be relevant for the understanding of the evolution of the inhomogeneities.

In this paper we focused on the formal understanding of the general solution, which we applied to two different setups.
For the case of two non-comoving dust components in a flat $\Lambda$CDM universe we found rich and interesting physics as, for instance, the gravitational interaction between the two fluids which makes one component follow the collapse of the other, or the evolution of the peculiar velocities which are increasing with respect to the background but decreasing between the two fluids.

Finally, for the case of clustering dark energy, we have been able to follow in an exact way the collapse of an overdensity to which also the dark energy contributes, and we confirmed that the evolution agrees with the theoretical growth rates during matter domination.
We also considered the case of a small speed of sound ($c_{s} \sim 10^{-5}$) and of a phantom equation of state, which also showed the expected dynamics.
These applications show some of the interesting features of this $n$-fluid component solution, which we will further investigate in forthcoming work, both from a theoretical and observational point of view.

\acknowledgments

It is a pleasure to thank Krzysztof Bolejko, Marie-No\"elle C\'el\'erier, Woei Chet Lim, Leandros Perivolaropoulos, Ignacy Sawicki and Wessel Valkenburg for useful comments and discussions.
M.~P.~acknowledges financial support from the Magnus Ehrnrooth Foundation.

\appendix

\section{Expansion tensor} \label{exten}

In this Appendix we will study the kinematical properties of a generic four-velocity field $u^{\alpha}$, which may be identified with the $i$:th fluid-component velocity $u_i^{\alpha}$ or with the reference-frame velocity $u_{\rm rf}^{\alpha}$.

The acceleration of $u_{\alpha}$ is given by $a_{\alpha} \equiv u^{\mu} \, \nabla_{\mu} u_{\alpha}$, whose components are:
\begin{equation}
   a_t =  -e^{\lambda}\gamma \, v_p \, a
\qquad \textrm{and} \qquad
   a_r =  \gamma \, {v_{p} \over v_{c}} \, a \,,
\end{equation}
where the acceleration scalar $a$ is
\begin{equation} \label{ascal}
 a = \sqrt{g^{\alpha\beta} a_{\alpha}a_{\beta}} =v_{p}^{-1} \Big(\Theta_T+  v_p^2 \,  \Theta_R \Big ) \,,
\end{equation}
where $\Theta_T$ and $\Theta_R$ are defined below.
The expansion tensor $\Theta_{\alpha\beta}$ is then:
\begin{equation}
   \Theta_{\alpha\beta} = h^{\mu}_{\;\;\alpha}h^{\nu}_{\;\;\beta} \, \nabla_{(\mu} u_{\nu)} =\nabla_{(\alpha} u_{\beta)} + a_{(\alpha} u_{\beta)} ,
\end{equation}
where $h_{\alpha\beta}=g_{\alpha\beta}+u_{\alpha}u_{\beta}$ is the projection tensor on the hypersurface orthogonal to $u_{\alpha}$. In the second equality we have used the fact that $u^{\alpha} \nabla_{\beta} u_{\alpha}=0$.                          
As we are considering a spherically-symmetric spacetime the velocity is irrotational, i.e., there is no vorticity:
\begin{equation}
\omega_{\alpha\beta} = h^{\mu}_{\;\;\alpha}h^{\nu}_{\;\;\beta} \,    \nabla_{[\mu} u_{\nu]} =0 \,.
\end{equation}
$h_{\alpha\beta}$ is often referred to as the ``spatial metric'' because it is the induced metric on the space slices, $h_{\alpha\beta}= h^{\mu}_{\;\;\alpha}h^{\nu}_{\;\;\beta} \, g_{\mu\nu}$.
Moreover, $h_{\alpha\beta}$ is the first fundamental form of the hypersurface and the expansion tensor is (minus) the extrinsic curvature, $K_{\alpha \beta} = - \Theta_{\alpha\beta}$, which is the second fundamental form of the hypersurface.

The expansion scalar can be obtained without calculating explicitly the components of the expansion tensor by remembering that it is:
\begin{equation} \label{expst}
\Theta = \nabla_{\alpha}  u^{\alpha}= {\partial_{\alpha} (J u^{\alpha} )  \over J} = \partial_{\alpha}u^{\alpha} + {d \ln J \over d\tau} \,,
\end{equation}
where $J= \sqrt{-g} = J_T \, J_R \, J_A$, $g$ is the determinant of the metric and we have defined the following quantities:
\begin{equation}
J_T=e^{\lambda}
\,, \quad
J_R= {Y'  \over \sqrt{1+2E}} 
\quad \textrm{and} \quad
J_A=  Y^{2} \sin \theta \,.
\end{equation}
In the last expression of Eq.~(\ref{expst}), the first term can be interpreted as the ``newtonian'' expansion scalar and the second as the contribution from the metric.
We then define the temporal, radial and angular expansion scalars:
\begin{eqnarray}
\Theta_T &=&  \partial_{t}u^{t} + {d \ln J_T \over d\tau} = e^{-\lambda} \dot \gamma +  \gamma v_{c} \, \lambda ' \,, \\
\Theta_R &=&  \partial_{r}u^{r} + {d \ln J_R \over d\tau} =(\gamma v_{c})' +  \gamma \bigg( H_R  +S_{R} \bigg)  \,,  \\
\Theta_A &=&  \partial_{\theta}u^{\theta}+\partial_{\phi}u^{\phi} + {d \ln J_A \over d\tau} =    \gamma \bigg( 2 H_A +2 S_{A} \bigg) \,,
\end{eqnarray}
respectively, where we defined the quantities:
\begin{equation}
S_{R}= v_{c} \, {Y'' \over Y'} - v_{c} {E' \over1+2E}
\qquad \textrm{and} \qquad
S_{A}= v_{c} \, {Y' \over Y} \,,
\end{equation}
which are defined analogously to $H_{R}$ and $H_{A}$ of Eqs.~(\ref{hubbleR}-\ref{hubbleA}) and can be understood as ``spatial'' expansion rates relevant for non-comoving fluids.
From Eq.~(\ref{ascal}) follows that the expansion scalar for a geodesic four-velocity field is $\Theta_{\rm geod.} = \Theta_R / \gamma^{2} +\Theta_A$.

In order to calculate the shear we need the non-vanishing components of $\Theta_{\alpha\beta}$:
\begin{eqnarray}
   \Theta_{tt}           & = & e^{2\lambda}\gamma^2 v_p^2 \, ( \Theta_T + \Theta_R)      \,, \qquad \quad
   \Theta_{rt}           = -e^{\lambda}\gamma^2\frac{v_p^2}{v_c}\,  ( \Theta_T + \Theta_R)    \,,    \\
   \Theta_{rr}           & = &  \gamma^2 {v_p^2 \over v_{c}^{2}} \,  ( \Theta_T + \Theta_R)  \,,  \qquad \quad
   \Theta_{\theta\theta} =   \Theta_{\phi\phi}/ \sin^2\theta = \frac{1}{2}Y^2 \, \Theta_A  \,.
\end{eqnarray}
Note that $\Theta_{tt}$ and $\Theta_{rt}$ are nonzero as we are using coordinates in general not comoving with $u^{\alpha}$.
The components of the expansion tensor clearly satisfy:
\begin{equation}
   \Theta = g^{\alpha\beta} \, \Theta_{\alpha \beta}  =\Theta_T + \Theta_R +\Theta_A  \,. \label{expascal}
\end{equation}
The shear tensor $\sigma_{\alpha\beta}$ is then defined in terms of $\Theta_{\alpha\beta}$ as
\begin{equation}
   \sigma_{\alpha\beta} = \Theta_{\alpha\beta} - \frac{1}{3}\Theta \, h_{\alpha\beta} \,,
\end{equation}
and its non-vanishing components are:
\begin{eqnarray}
   \sigma_{tt}           & = &  \frac{2}{\sqrt{3}}e^{2\lambda}\gamma^2v_p^2 \,  \sigma    \,,  \qquad \quad
   \sigma_{rt}           =  -\frac{2}{\sqrt{3}}e^{\lambda}\gamma^2 \frac{v_p^2}{v_c} \,  \sigma    \,, \\
   \sigma_{rr}           & = &  \frac{2}{\sqrt{3}}\gamma^2\frac{v_p^2}{v_c^2} \,  \sigma    \,,\qquad \quad
   \sigma_{\theta\theta} =  { \sigma_{\phi\phi} \over \sin^{2}\theta} =  - \frac{1}{\sqrt{3}}Y^2 \,  \sigma    \,,
\end{eqnarray}
where the invariant $\sigma$ of the shear tensor is given by:
\begin{eqnarray}
   \sigma^2 = {1 \over 2} g^{\alpha\mu}  g^{\nu\beta}  \sigma_{\mu\nu}\sigma_{\alpha\beta} = \frac{1}{3} \left( \Theta_T + \Theta_R - \frac{1}{2}  \Theta_A     \right)^2  \,.
\end{eqnarray}

We may now particularize the above expressions for the case of the reference-frame velocity $u_{\rm rf}^{\alpha}$:
\begin{eqnarray}
a_{\rm rf} &=&  {v_{c} \over v_{p}} \, \lambda'  \,, \qquad
a_{r,\, \rm rf}=\lambda'  \,, \qquad
a_{t,\, \rm rf}=0  \,, \label{exparf1} \\
\Theta_{\rm rf} &=&  \Theta_{R,\,\rm rf} +\Theta_{A,\,\rm rf} = H_R+ 2 H_A , \qquad
\sqrt{3} \sigma_{\rm rf} = \Theta_{R,\,\rm rf} - \frac{1}{2} \Theta_{A,\,\rm rf}  = H_R -  H_A .
\end{eqnarray}
Note that, as $\Theta_{T}$ vanishes, $\lambda$ appears in the expansion scalar and shear only through the derivatives with respect to the proper time $d/d\tau_{\rm rf}=u_{\rm rf}^{\alpha} \, \partial_{\alpha}= e^{-\lambda} \partial/\partial t$. This shows that a non-geodesic reference frame only changes the proper time with which the expansion rate is measured, but not its functional form (different is the case for the acceleration rates of Eqs.~(\ref{accR}-\ref{accA})).
Moreover, these last results show that the definitions in Eqs.~(\ref{hubbleR}-\ref{hubbleA}) are relative to the reference-frame velocity.

\section{The Lema\^{i}tre metric} \label{pLTB} 

It is useful to consider the case of only one fluid component, which could be seen as the Lema\^{i}tre metric \cite{Lemaitre:1933gd} in a possibly non-comoving reference frame.
We will discuss some properties of this solution, which may help us better understand the multi-fluid non-comoving case.
We will not discuss the dynamical evolution of the Lema\^{i}tre metric, which can be easily deduced from the set of Eqs.~(\ref{fset1}-\ref{fset5}).

Let us start by considering the Eqs.~(\ref{BAMd}-\ref{BAMp}) for $\dot F$ and $F'$.
By taking the combinations giving the derivative of $F$ with respect to $\tau$ and $\sigma$ it is possible to obtain the following simpler expressions:
\begin{eqnarray}
{d F  \over d \tau }& = & - p \, {d V_{\rm e} \over d\tau} =-4 \pi Y^{3} \, p  \, {\Theta_{A} \over 2}  \,, \label{Fti} \\
{d F  \over d \sigma } &=&   \rho \, {d V_{\rm e} \over d\sigma}   \,, \label{Fsi} 
\end{eqnarray}
where the Euclidean and actual volumes of the space slices are, respectively: 
\begin{equation}
V_{\rm e} =4 \pi \int_{0}^{r}  Y^{2} Y' d \hat r = {4\pi \over 3} Y^{3} \,, \qquad \quad
V = 4 \pi \int_{0}^{r}  {Y^{2} Y' \over \sqrt{1+2E}}  d \hat r \,.
\end{equation}
Note that Eqs.~(\ref{Fti}-\ref{Fsi}) give the gravitating mass of a non-comoving fluid.
In the fluid rest frame (the actual Lema\^{i}tre metric) the previous equations have the familiar form:
\begin{eqnarray}
\dot F &= & -4 \pi Y^{2} \dot Y\, p =  - p \,  \dot V_{\rm e} \,, \label{Ftir} \\
F' &=& 4 \pi Y^{2} Y'   \, \rho   =   \rho \,   V_{\rm e}' \,, \label{Fsir} 
\end{eqnarray}
so that the first gives the time evolution of $F$, which is constant only for vanishing pressure, and the second links the local density to the integrated gravitating mass.
Eqs.~(\ref{Ftir}-\ref{Fsir}) can be combined into the differential expression:
\begin{equation} \label{fpt1}
d F =  \rho \, dV_{e}|_{t={\rm const}} - p \, dV_{e}|_{r={\rm const}}  \,,
\end{equation}
which has the usual thermodynamical interpretation for a fluid in equilibrium (at constant entropy): for an increase $dV$ of volume at constant $t$ the total mass-energy is increased by the mass-energy density of the volume $dV$, and for an increase $dV$ at constant $r$ the total mass-energy is decreased by the work of the pressure against $dV$.
In this picture the coordinate $r$ has to be seen as a label for the mass-energy density.

It is interesting to point out that the previous equations use the Euclidean volume element and so the gravitating (or Euclidean) mass $F$ does not coincide with the invariant mass $M$ which is related to the local density by~\cite{Bondi:1947av}:
\begin{equation} \label{im1}
M' = 4 \pi{Y^{2} Y' \over \sqrt{1+2E}}   \; \rho = \rho \, V' \,.
\end{equation}
We can then evaluate the time derivative of $M$ at constant $r$:
\begin{eqnarray}
\dot M &=& 4 \pi \int_{0}^{r}  {Y^{2} Y' \over \sqrt{1+2E}} \, \rho   { \left( J_{R} J_{A} \rho \right)\dot{}  \over J_{R} J_{A} \rho} \, d \hat r  
= 4 \pi \int_{0}^{r}  {Y^{2} Y' \over \sqrt{1+2E}} \, \rho  \left( { \dot \rho \over \rho}  + e^{\lambda} \Theta_{\rm rf} \right) \, d \hat r \nonumber \\
&=&- 4 \pi  \int_{0}^{r}  {Y^{2} Y' \over \sqrt{1+2E}}  e^{\lambda} \Theta_{\rm rf} \, p  \, d \hat r = -  p \, \dot V \,,  \label{dM1}
\end{eqnarray}
where to go from the first to the second line we have used the conservation equation (\ref{fset4}) and for the last equality we have assumed a homogeneous pressure $p'=0$.
Note that if $p' = 0$, then from Eqs.~(\ref{fset5}) and (\ref{fset2}) follows that $\dot E = 0$, even if $M$ and $F$ are evolving.
Similarly to Eq.~(\ref{fpt1}), we can combine Eqs.~(\ref{im1}-\ref{dM1}) into:
\begin{equation} \label{fpt2}
d M =  \rho \, dV|_{t={\rm const}} - p \, dV|_{r={\rm const}}  \,.
\end{equation}
Note, however, that in this last equation not only a different volume is used with respect to Eq.~(\ref{fpt1}), but that Eq.~(\ref{fpt2}) is not valid for $p' \neq 0$, which implies a time-dependent curvature (that can be interpreted as total energy per unit of mass).
In this latter case the usual thermodynamical expression, where only the pressure at the boundary matters, does not seem to hold and the value of the pressure at any $r$ matters as shown by the last integral in Eq.~(\ref{dM1}).

Finally, it is interesting to rewrite Eq.~(\ref{fset1}) as:
\begin{equation} \label{eucdy}
H_A^2 = \frac{2 G F}{Y^3}+ \frac{2 E}{Y^2} = {\kappa \over 3}  \langle \rho \rangle_{\rm e}  -   \left \langle {{\cal{R}} \over 6} \right \rangle_{\rm e} \,,
\end{equation}
where the Euclidean average for the generic quantity $X$ is defined as:
\begin{equation} \label{euave}
 \langle X \rangle_{\rm e}  = {4 \pi \int_{0}^{r}  Y^{2} Y'   \, X \, d \hat r \over V_{\rm e}} \,,
\end{equation}
and ${\cal{R}}$ is the spatial Ricci scalar of the rest frame, which is the trace of the Ricci tensor of the metric $g_{ij}$ on the hypersurface of constant $t$:
\begin{equation}
{\cal{R}} = - {4 (EY)' \over Y^2Y'} \,.
\end{equation}
The factor of $1/6$ in Eq.~(\ref{eucdy}) can be understood by the fact that for the FLRW model it is ${\cal{R}} /6 = k/a^{2}$.
From Eq.~(\ref{eucdy}) follows that the expansion rate is sourced by the Euclidean average of the local density and curvature, and not by the actual averages.
This is a potential source for ``strong'' backreaction effects \cite{Kolb:2008bn,Kolb:2009rp} of the inhomogeneities on the evolution of the background (see, for example, Appendix B of \cite{Marra:2011ct} and also \cite{Mattsson:2010vq,Sussman:2011na}).

\section{The \boldmath $W_{n}$ and \boldmath $\delta_{n}$ functions} \label{wdn}

The $W_{n}(x, \alpha)$ step functions introduced in \cite{Valkenburg:2011tm} are $C^{n}$ everywhere and go from 1 to 0 for $x$ from 0 to 1 with a transition which is steeper for higher $\alpha \in [0,1[$.
In this paper we used $W_{1}$ and $W_{3}$, which have the following explicit form:
\begin{equation*} 
W_{1}(x, \alpha)= \left\{
  \begin{array}{ll}
    1  &  0\le x < \alpha \\
    \frac{1}{2} + \frac{1}{2}  \sin \left[ \pi \left(\frac{1}{2} - \frac{x - \alpha}{1-\alpha} \right) \right]  &  \alpha \le x < 1  \\
    0 &   1 \le x
  \end{array}\right. \,,
\end{equation*}
and
\begin{equation*} 
W_{3}(x, \alpha)= \left\{
  \begin{array}{ll}
    1  &  0\le x < \alpha \\
    \frac{1}{4\pi^2}\left \{1 + \pi^2 \left [ 4 - 8 \left(\frac{x - \alpha}{1-\alpha}\right)^2 \right] -  \cos \left( 4\pi \frac{x - \alpha}{1-\alpha} \right) \right \}  &  \alpha \le x < \frac{1+\alpha}{2} \\
    \frac{1}{4\pi^2}\left[ - 1 + 8\pi^2 \left (\frac{x - \alpha}{1-\alpha} -1 \right)^2 +  \cos \left( 4\pi\frac{x - \alpha}{1-\alpha} \right) \right] &  \frac{1+\alpha}{2} \le x < 1 \\
    0 &   1 \le x
  \end{array}\right. 
\end{equation*}
From the $W_{n}$ function we then built the following general contrast $\delta_{n}$:
\begin{equation} \label{rhoex} 
\delta_{n}(r,r_{t},\Delta_{1,2},\alpha_{1,2})  = \left\{
  \begin{array}{ll}
    \Delta_{2} +  \Delta_{1} \, W_{n}\bigg({r \over r_{t}}, \alpha_{1} \bigg)  &  0 \le r < r_{t} \\
    \Delta_{2} \, W_{n}\bigg({r-r_{t} \over r_{b}-r_{t}}, \alpha_{2}\bigg)  &  r_{t} \le r < r_{b} \\
    0 &   r_{b} \le r
  \end{array}\right. \;,
\end{equation}
where $\Delta=\Delta_{1}+\Delta_{2}$ is the contrast between the center ($r=0$) of the spherical inhomogeneity and the border ($r=r_{b}$), and $\Delta_{2}$ is the contrast at $r_{t}$.
We have used $\delta_{n}$ to model inhomogeneities in $\rho_{M}$, $F_{N}$ and $v_{c}$ in Section \ref{colla}.


\end{document}